\documentclass[preprintnumbers,amsmath,amssymb, prd, nofootinbib]{revtex4}
\usepackage{amsmath,amssymb,graphicx}
\usepackage {amssymb}
\usepackage{color}
\newcommand{\nc}{\newcommand}
\nc{\ba}{\begin{eqnarray}}
\nc{\ea}{\end{eqnarray}}
\newcommand\be{\begin{equation}}
\newcommand\ee{\end{equation}}

\nc{\bfk}{{\mathbf{k}}}
\nc{\bfq}{{\mathbf{q}}}
\nc{\bfp}{{\mathbf{p}}}
\nc{\e}{{\bf{e}}}
\nc{\calR}{{\cal R}}
\nc{\calP}{{\cal P}}

\input{epsf.sty}

\begin{document}


\title{Primordial Anisotropies in Gauged Hybrid Inflation}

\author{Ali Akbar Abolhasani$^{1}$}
\email{abolhasani-AT-ipm.ir}
\author{Razieh Emami$^{1, 2}$}
\email{emami-AT-ipm.ir}
\author{Hassan Firouzjahi$^{3}$}
\email{firouz-AT-mail.ipm.ir}

\affiliation{$^1$School of Physics, Institute for Research in
Fundamental Sciences (IPM),
P.~O.~Box 19395-5531,
Tehran, Iran}

\affiliation{$^2$ Abdus Salam International Centre for Theoretical Physics,  Strada Costiera 11, 34151, Trieste, Italy}

\affiliation{$^3$School of Astronomy, Institute for Research in
Fundamental Sciences (IPM),
P.~O.~Box 19395-5531,
Tehran, Iran}

\begin{abstract}
\vspace{0.3cm}

We study primordial anisotropies generated in the model of gauged hybrid inflation in which the complex waterfall field is charged under a $U(1)$ gauge field.
Primordial anisotropies are generated either actively during inflation or from inhomogeneities  modulating the surface of end of inflation during waterfall transition.
We present a consistent $\delta N$ mechanism to calculate the anisotropic power spectrum and bispectrum.
We show that the primordial anisotropies generated at the surface of end of inflation do not depend on the number of e-folds  and therefore do not produce dangerously large anisotropies associated with the IR modes. Furthermore, one can find the parameter space that the anisotropies generated from the surface of end of inflation cancel the anisotropies generated during inflation, therefore relaxing the  constrains on model parameters imposed from
IR anisotropies.  We also show that the gauge field fluctuations induce a red-tilted power spectrum so the  averaged power spectrum from the gauge field can change the total power spectrum from blue to red. Therefore, hybrid inflation, once gauged under a $U(1)$ field, can be consistent with the cosmological observations.

\vspace{0.3cm}

\end{abstract}

\date\today

\maketitle

\section{Introduction}

Precision observations from  WMAP  \cite{Bennett:2012zja,Hinshaw:2012fq} and PLANCK   \cite{Ade:2013lta, Ade:2013uln} have provided strong supports for inflation as the leading theory for early universe and as a mechanism to generate the primordial seeds for the  cosmological structures.  The basics predictions of inflation are near scale-invariant, near Gaussian and near adiabatic perturbations on Cosmic Microwave Background (CMB) which are well consistent with the recent cosmological observations. The simplest  models of inflation consist  a single scalar field which is minimally coupled to gravity with an appropriate flat potential to support a long enough period of slow-roll inflation.

Statistical isotropy of universe is a key assumption in standard cosmology which is well-motivated theoretically and is supported by different cosmological observations at different scales and red-shifts. Statistical isotropy on large scale is also motivated from the Copernicus point of view that there may not exist a preferred direction or reference point in sky. Having this said, there are evidences for the violation of statistical isotropy both in WMAP results and also in PLNCK results, for a detailed review see \cite{Ade:2013nlj}. The PLANCK
data implies a detection of dipole asymmetry.

The possibility of a quadrupolar asymmetry in primordial curvature perturbation power spectrum was also studied by PLANCK team. If one parameterizes the primordial power spectrum via
\ba
\label{g-def}
\calP_\calR = \calP_{\calR }^{(0)} \left( 1+ g_* (\hat {\bf p}. \hat{\bf k})^2 \right)
\ea
in which $\calP_{\calR }^{(0)}$ represents the isotropic power spectrum,
$\hat{\bf k}$ represents the momentum direction in Fourier space and $\hat {\bf p}$
is the preferred direction in the sky, then the PLANCK data shows $ -0.05< g_* < 0.05$ and
$-0.36 < g_* < 0.36$ from $L=0, L=2$ modes respectively at 95 \% CL. This was also studied recently in \cite{Komatsu} using the PLANCK data yielding $g_* =0.002 \pm 0.016$ at 68\% CL.  Although there is no observational indication for quadrupolar asymmetry in primordial power spectrum, but the the possibility of having a primordial quadrupolar asymmetry is intriguing both theoretically and observationally.

An interesting mechanism in generating quadrupolar asymmetry in primordial power spectrum
is to consider anisotropic inflation in the presence of a $U(1)$ gauge field. As is well-known, the simple $U(1)$ gauge field minimally coupled to gravity suffers from the conformal invariance such that both at the level of background and also at the level of perturbations the gauge fields are exponentially diluted. In order to overcome this difficulty, it is necessary to break the conformal invariance during inflation. A simple mechanism to break conformal invariance  is to couple the gauge field  to the inflaton field $\phi$ non-minimally via
$f(\phi)^2 F_{\mu}F^{\mu}/4$ in which $F_{\mu \nu}$ is the gauge field strength and $f(\phi)$ is the  inverse of the gauge kinetic coupling \cite{Turner:1987bw, Ratra:1991bn}.
With an appropriate choice of the coupling $f(\phi)$ one can break the conformal invariance such that an attractor mechanism is generated for the gauge field dynamics \cite{Watanabe:2009ct}.
The key property of this attractor mechanism is that the fraction of the
gauge field energy density to the total energy density reaches at the order of slow-roll parameters and remains nearly constant once the attractor phase has been reached \cite{Watanabe:2009ct}.  As a result, one can generate statistical anisotropy which is small but nonetheless observable.  There have been many works on anisotropic inflation and magneto-genesis with the conformal coupling $f(\phi)^2 F_{\mu}F^{\mu}$ \cite{Demozzi:2009fu, Martin:2007ue, Emami:2009vd, Emami:2010rm, Emami:2011yi, Kanno:2009ei, Caldwell:2011ra, Jain:2012vm,  Watanabe:2009ct, Kanno:2010nr, Murata:2011wv, Bhowmick:2011em, Hervik:2011xm, Thorsrud:2012mu, Dimopoulos:2010xq, Yamamoto:2012tq, Moniz:2010cm, Boehmer:2007ut, Koivisto:2008xf, Maleknejad:2011jr, Maleknejad:2012as, Yokoyama:2008xw,  Lyth:2012vn,    Baghram:2013lxa,
Dimopoulos:2009vu, Dimopoulos:2012av, Dimastrogiovanni:2010sm, ValenzuelaToledo:2009af, Himmetoglu:2008zp, Giovannini:2001nh, Giovannini:2007rh, Kunze:2013hy,  Kahniashvili:2012vt, Rodriguez:2013cj, BeltranAlmeida:2011db, Urban:2013spa, Ohashi:2013qba, Ohashi:2013mka, Ohashi:2013pca, Fujita:2013qxa, Shiraishi:2013sv, Linde:2012bt, Bartolo:2013msa}, for a review see \cite{Kandus:2010nw, Soda:2012zm, Maleknejad:2012fw}.

The idea of generating primordial anisotropy at the surface of end of inflation was pioneered
by Yokoyama and Soda
in \cite{Yokoyama:2008xw}. In this scenario, the surface of end of inflation is controlled by the waterfall field which is coupled to a $U(1)$ gauge field. It is argued in \cite{Yokoyama:2008xw} that the inhomogeneities sourced by the gauge field fluctuations $\delta A_\mu$ at the surface of end of inflation can lead to statistical anisotropies. In the spirit this is similar to idea used in   \cite{Lyth:2005qk} to generate curvature perturbations at the end of inflation for models containing interacting scalar fields. The proposal employed in \cite{Yokoyama:2008xw} was reviewed  critically in  \cite{Emami:2011yi} using a more rigorous application of $\delta N$ formalism. It is argued in \cite{Emami:2011yi} that the mechanism as suggested in  \cite{Yokoyama:2008xw} does not work since the evolution of the gauge field during inflation has not been taken into account.

One important development in this context
was the extension of $\delta N$ and separate universe approach \cite{Sasaki:1995aw, Wands:2000dp, Lyth:2004gb, Lyth:2005fi, Naruko:2012fe, Naruko:2012um, Sugiyama:2012tj}
to anisotropic backgrounds such as Bianchi I universe \cite{Abolhasani:2013zya}. In  \cite{Abolhasani:2013zya} the $\delta N$ formalism for anisotropic inflation scenarios is presented in which the back-reactions of the gauge field on inflaton field in the attractor regime and the evolution of the gauge field dynamics during inflation has been considered. In particular, it is shown that in anisotropic inflation model  of \cite{Watanabe:2009ct}, the gauge field fluctuations enter into
$\delta N$ formula via $\delta \dot A/\dot A$ so the fraction  $\delta \dot A/\dot A$ behaves like the ratio $\delta \phi/\phi$ for a light scalar field in a dS background. For earlier works on
$\delta N$ in models with vector fields see  \cite{Dimopoulos:2008yv, ValenzuelaToledo:2011fj}.

In this work, we study the model of gauged hybrid inflation in which the waterfall is a complex scalar field coupled to the $U(1)$ gauge field. The interesting property of this model is that  primordial anisotropies can be generated both during inflation and also at the surface of end of inflation via waterfall phase transition. Similar to  \cite{Abolhasani:2013zya}, we present the  $\delta N$ formalism to calculate the anisotropic power spectrum and the bispectrum. This also allows us to revisit the original idea of Yokoyama-Soda in \cite{Yokoyama:2008xw}. One novel feature of generating anisotropies at the surface of end of inflation is that the large unwanted anisotropy sourced by the cumulative IR contributions of super-horizon modes \cite{Bartolo:2012sd}, scaling like $N^2$ as a function of the total number of e-folds $N$, does not show up.

The rest of the paper is organized as follow. In Section \ref{hybrid} we present our set up, the gauged hybrid inflation model, in which both the inflaton field and the gauge field are  coupled to the waterfall field responsible to terminate inflation abruptly. Towards the end of this Section we  we present our $\delta N$ formula to second order in inflation and gauge field perturbations. In Section \ref{power-spec} we present the power spectrum analysis and calculate $g_*$ in this model. In Section \ref{bi-spec} we present the bispectrum analysis followed by our discussions and summaries in Section \ref{summary}. We relegates some technicalities of the $\delta N$ analysis into Appendices.

\section{Gauged Hybrid Inflation}
\label{hybrid}

In this section we study our setup of anisotropic inflation, the gauged hybrid inflation model.
The system contains the inflaton field $\phi$, the complex waterfall field $\psi$ and the gauge field $A_\mu$. As in standard hybrid inflation scenarios \cite{Linde:1993cn, Copeland:1994vg},  the waterfall field is responsible to terminate inflation abruptly. In this model, both inflaton and the gauge field are coupled to waterfall field so the surface of end of inflation is controlled both by the inflaton field and the gauge field. Similar to the mechanism of  generating curvature perturbations from
inhomogeneities at the surface of end of inflation \cite{Lyth:2005qk, Bernardeau:2004zz, Dvali:2003em, Assadullahi:2012yi},  the gauge field fluctuations source anisotropy at the surface of end of inflation as advocated in  \cite{Yokoyama:2008xw}. Here we study anisotropies generated at the surface of end of inflation as well as anisotropies generated  actively during inflation  \cite{Dulaney:2010sq, Gumrukcuoglu:2010yc, Watanabe:2010fh, Funakoshi:2012ym, Yamamoto:2012sq, Emami:2013bk}
using a consistent $\delta N$ formalism following the method employed in \cite{Abolhasani:2013zya}.


\subsection{The model}

The model of gauged hybrid inflation is based on the action \cite{Emami:2010rm, Emami:2011yi}
\ba
\label{action3} S=\int d^4 x  \sqrt{-g} \left [ \frac{M_P^2}{2} R - \frac{1}{2} \partial_\mu \phi
\,  ^\mu\phi- \frac{1}{2} D_\mu \psi
\,  D^\mu\bar\psi - \frac{f^{2}(\phi)}{4} F_{\mu \nu} F^{\mu
\nu}- V(\phi, \psi, \bar \psi) \right]  \, ,
\ea
in which  $\phi$ is the inflaton field,  $\psi$ is the complex waterfall field and the  covariant derivative is given by
\ba
D_\mu \psi = \partial_\mu  \psi + i \e \,  \psi  \, A_\mu \, .
\ea
Here $\e$ is the  gauge coupling representing  the interaction between the gauge field and the waterfall field.  Finally,  the gauge field strength is defined  by
\ba F_{\mu \nu} = \nabla_\mu A_\nu
- \nabla_\nu A_\mu  = \partial_\mu A_\nu - \partial_\nu A_\mu \, .
\ea
As mentioned before, the gauge kinetic coupling $f(\phi)$ is chosen appropriately to break the conformal invariance such that the gauge field survives the expansion both at the background level as well as at the perturbations level.

As usual we consider the  configurations in which the potential is axially symmetric and $V(\psi, \bar \psi , \phi)= V(\chi, \phi)$ in which the waterfall field is  decomposed into the radial and the angular parts  $ \psi(x) = \chi(x) \,  e^{i \theta(x)}$. The potential is given by the  standard hybrid inflation potential \cite{Linde:1993cn}
\ba
\label{pot3} V(\phi, \chi)=\frac{\lambda}{4 }  \left(  \chi^2 - \frac{M^2}{\lambda} \right)^2 + \frac{g^2}{2} \phi^2 \chi^2 + \frac{m^2}{2} \phi^2 \, ,
\ea
in which $\lambda$ and $g$ are two dimensionless couplings and $m$ and $M$ are two mass parameters.

In our analysis below we choose the unitary gauge in which $\theta(x) =0$ so we neglect the phase
of the waterfall in the  background and perturbations analysis.

To be specific, we assume that the background gauge field is turned on along the $x$ direction with the background component $A_{\mu}=(0,A(t),0,0)$. In the presence of the gauge field we lose the rotational invariance and the system reduces to Bianchi type I universe. As studied in \cite{Watanabe:2009ct} with the appropriate choice of the conformal coupling $f(\phi)$ the system reaches the attractor regime in which the fraction of the gauge field energy density to the total energy density becomes at the order of slow-roll parameter. In particular, the back-reactions of the gauge field to inflaton dynamics is non-negligible. As we shall see this play crucial roles in calculating the anisotropies generated in the system.

The background Bianchi I metric is given by
\ba
\label{metric}
ds^2 &=& - dt^2 + e^{2\alpha(t)}\left( e^{-4\sigma(t)}d x^2 +e^{2\sigma(t)}(d y^2 +d z^2) \right)  \nonumber\\
&\equiv&   - dt^2 + a(t)^2 dx^2 + b(t)^2 (dy^2 + dz^2)
\ea
in which  $\alpha(t)$ represents  the average number of e-foldings, $\dot \alpha$
represents the corresponding isotropic Hubble expansion rate while $\dot \sigma(t)$
represents the anisotropic expansion rate. Note that because of our assumption that the gauge field is turned on along the $x$-direction, we still have a subset of two-dimensional rotational symmetry in $y-z$ plane.  Demanding that the level of anisotropy is small in order to be consistent with the cosmological observations we require
$|\dot \sigma/\dot \alpha| \ll 1$.

The total energy density from the gauge field and the inflaton field is
\ba
\label{energy}
{\cal E}=V(\phi,\chi) + e^{-2 N + 4 \sigma} \left( \frac{1}{2}f^2(\phi) \dot
A^2+\frac{\mathbf{e}^2\chi^2}{2} A^2 \right) \, ,
\ea
in which we have neglected the scalar fields kinetic energy in the slow-roll limit.
Also to simplify the notation, we have used $\alpha(t) =N(t)$ representing the number of e-foldings.
From Eq. (\ref{energy})  we see that the gauge field has two contributions in total energy density, the first term in the big bracket coming from its kinetic energy and the second term in the big bracket coming from its contribution to the potential energy with the coupling $\e^2 \chi^2$.  As we shall see  both terms play key roles in generating anisotropies. The latter contribution has the interesting new effect  that the gauge coupling $\mathbf{e}$ induces a new time-dependent mass term $\mathbf{e}^2 e^{-2 N} A_\mu A^\mu $
for the waterfall field.  As in standard hybrid inflation \cite{Linde:1993cn, Copeland:1994vg}
we consider the vacuum dominated regime in which $\chi$ is very heavy during inflation so
it rapidly goes to its instantaneous minimum $\chi=0$ during inflation. In standard hybrid inflation  models  inflation abruptly ends when the inflaton field reaches the critical value, $\phi=\phi_c \equiv \frac{M}{g}$ in which the waterfall field becomes tachyonic and rolls down quickly to its global minimum $|\chi |=\mu\equiv M/\sqrt{\lambda}, \phi=0$. However, in our system  the  coupling of gauge field to waterfall field modifies the  surface of end of inflation. Specifically, the effective mass of waterfall is
\ba
\label{chi-mass}
\frac{\partial^2 {\cal E}}{\partial \chi^2}\large|_{\chi=0}  = g^2 (\phi^2 - \phi_c^2) + \mathbf{e}^2 e^{- 2 N+4\sigma} A^2 \, .
\ea
 In the absence of the gauge field, the moment  of waterfall instability is determined  when $\phi= \phi_c$. However, in the presence of gauge field the condition of waterfall instability  is modified. More specifically, the condition of waterfall phase transition from  Eq. (\ref{chi-mass})  can be rewritten as
\ba
\label{transition}
 \phi_f^2 + \frac{\mathbf{e}^2}{g^2} A_f^2  =\phi_c^2
\ea
in which $\phi_f$ and $A_f$ represents the fields values at the surface of end of inflation.
Note that  in the absence of gauge field the waterfall phase transition happens at
$ \phi_f = \phi_c = M/g$.

Note that we have chosen the convention such that the time of end of inflation corresponds to $N\equiv N_f=0$ and count the number of e-foldings such that  $N(t)= -\int_t ^{t_f} dt H <0$.  In order to solve the flatness and the horizon problem in FRW cosmology we need at least 60 e-foldings so  at the start of inflation   $N\equiv N_i \simeq -60$.

As mentioned before, we consider the limit that the waterfall field is very heavy during inflation and it quickly rolls down to its instantaneous minimum $\chi=0$ so the potential driving inflation approximately is
\ba
\label{V-app}
V\simeq \frac{M^4}{4\lambda}+\frac{1}{2}m^2\phi^2  \, .
\ea
Imposing the slow-roll condition that the  inflaton field is light during inflation requires $p_c\gg1$ in which $p_c$ is given via
\ba
\label{pc-def}
p_c \equiv \frac{M^4}{2 \lambda m^2 M_P^2} \, .
\ea
On the other hand, the assumption that the waterfall field is much heavier than the Hubble expansion rate during  inflation requires
$\lambda M_P^2 /M^2 \gg1$. Furthermore, the condition of vacuum domination during inflation implies $\lambda/g^2 \ll M^2/m^2$. Finally, we work in the limit where the waterfall phase transition is very sharp and inflation abruptly ends once the tachyonic instability occurs, corresponding to  $\lambda M_P^2 /M^2 \gg p_c$ \cite{Abolhasani:2010kr}.


\subsection{The attractor Solution}
Here we study the attractor solution in the light of \cite{Watanabe:2009ct}.
The background fields equations are given by
\ba
\label{back-A-eq3}
\partial_t{\left(  f^2(\phi) e^{\alpha + 4 \sigma} \dot A \right)}& =& - \e^2 \chi^2 e^{\alpha + 4 \sigma}  A \\
\label{back-phi-eq3}
\ddot\phi+3\dot \alpha\dot \phi+ \phi(m^2+g^2\chi^2) -f(\phi)f_\phi(\phi)\dot A^2 e^{-2\alpha+4\sigma}&=&0  \\
\label{back-chi-eq3}
\ddot\chi+3\dot \alpha\dot \chi+ \left(\lambda(\chi^2-\frac{M^2}{\lambda})+g^2\phi^2 \right)\chi
+\e^2\chi A^2 e^{-2\alpha+4\sigma}&=&0  \\
\label{Ein1-eq3}
\frac{1}{2}\dot
\phi^2+\frac{1}{2}\dot \chi^2+V(\phi,\chi)+ \left( \frac{1}{2}f^2(\phi)\dot
A^2 +\frac{\e^2\chi^2}{2}A^2 \right) e^{-2\alpha+4\sigma}
&=&
3 M_P^2 \left(\dot \alpha^2-\dot \sigma^2 \right)  \\
\label{Ein2-eq3}
V(\phi,\chi)+  \left(  \frac{1}{6}f^2(\phi)\dot
A^2+\frac{\e^2\chi^2}{3}A^2  \right)e^{-2\alpha+4\sigma}
&=& M_P^2 \left( \ddot \alpha    + 3 \dot \alpha^2 \right)  \\
\label{anisotropy-eq3}
\left(\frac{1}{3}f^2(\phi)\dot A^2  -\frac{\e^2\chi^2}{3}A^2 \right) e^{-2\alpha+4\sigma}
&=& M_P^2\left( 3\dot \alpha \dot \sigma+ \ddot \sigma\right)\, .
\ea
in which a dot indicates derivative with respect to $t$.
As can be seen in the above equations of motion, this system of coupled differential equations is too complicated to be solved analytically. However, if we note that we are looking in small anisotropy limit,
$|\dot \sigma/\dot \alpha| \ll 1$, the situation simplifies considerably. In this limit, one can neglect the contribution of the gauge field in Friedmann equation, Eq. (\ref{Ein1-eq3}), so the background expansion is controlled by the isotropic potential term as in usual isotropic slow-roll inflationary scenarios. However, the effects of the gauge field  becomes important in inflaton dynamics so one can not neglect its back-reaction in Klein-Gordon equation, Eq. (\ref{back-phi-eq3}). Fortunately, we can solve the system in this limit as we shall see below.

Note that since  the gauge coupling $\e$ plays role only near the end of inflation, one can safely neglect the right hand side of Eq. (\ref{back-A-eq3}) for the most period of inflation. Therefore, one can easily solve Eq. (\ref{back-A-eq3}) to obtain
\ba
\label{gaugefield}
\dot{A_{x}}= f(\phi)^{-2}e^{-\alpha(t)-4\sigma(t)}p_{A} \, ,
\ea
where $p_{A}$ is a constant of integration.

We are interested in the small anisotropy limit, $|\dot \sigma/H | \ll 1 $, so the background expansion is mainly supported by the isotropic potential term as in conventional models of inflation.  In order
for the anisotropy to be small, we demand that $R \ll 1$ in which
\ba
\label{R-def}
R \equiv \frac{\dot A^2 f(\phi)^2 e^{-2 \alpha}}{2 V} \, .
\ea
In this view $R$ measures the ratio of the electric field energy density, $\rho_{em}$, associated with the gauge field  to the total potential energy density. Therefore, to have  small anisotropies, we require $\rho_{em} \ll V$.
Although the anisotropy is small, $R\ll 1$, so the Hubble expansion rate in modified Friedmann equation (\ref{Ein1-eq3}) is mainly dominated by the isotropic potential term, but the back-reactions of the gauge field on the inflaton field induce an effective mass for the inflaton  as given by the last term in Eq. (\ref{back-phi-eq3}). This in turn will affect the dynamics of the inflaton field.

Now, let us determine the form of the conformal coupling
$f(\phi)$ necessary to break the conformal invariance and to obtain a near scale-invariant gauge field power spectrum. As studied in \cite{Watanabe:2009ct} during the attractor limit we require $R$ to be small but nearly constant. For this to happen we require
$f(\phi) \propto  e^{-2c \alpha} = a^{-2c} $ with $c>1$ is a numerical parameter of the model.   Indeed, for a generic potential $V(\phi)$
the background scale factor is given by
\ba
\label{a-scale}
a \propto \exp \left[ - \int d \phi \frac{V}{ M_P^2V_\phi} \right] \, .
\ea
So if we choose
\ba
\label{f-scale}
f \propto \exp \left[ 2 c \int d \phi \frac{V}{M_P^2V_\phi} \right]
\ea
we obtain  $f \propto a^{-2c}$ as required. Therefore, as mentioned in \cite{Watanabe:2009ct},  the exact form of $f$ therefore depends on the  form of the potential. For the hybrid potential given approximately by Eq. (\ref{V-app}) one obtains
\ba
\label{N-int}
\int d \phi \frac{V}{ M_P^2V_\phi} = \frac{M^4}{4 \lambda M_P^2 m^2} \ln\left( \frac{\phi}{\phi_f} \right) = \frac{p_c}{2} \ln\left( \frac{\phi}{\phi_f} \right) \, ,
\ea
in which $p_c$ is defined in Eq. (\ref{pc-def}) and $\phi_f$ is the value of the inflaton field at the end of inflation $\eta = \eta_f$.
Therefore, to obtain the attractor solution,
from Eqs. (\ref{f-scale})  and (\ref{N-int}) we require
\ba
\label{f-eq}
f(\phi) = \left( \frac{\phi}{\phi_f} \right)^{c \, p_c} =   \left( \frac{\phi}{\phi_f} \right)^{p}
\ea
in which we have defined $p\equiv  c\,  p_c$. Alternatively, we can also write $f(\phi)$ in terms
of $a(\eta)$. Using Eq. (\ref{a-scale}), we get
\ba
\label{f-a-scale}
f(\phi) =  \left(  \frac{a (\eta)}{a_f} \right)^{-2 c } =  \left(\frac{\eta}{\eta_{f}} \right)^{2c} \, .
\ea
In order for the gauge field fluctuations to survive the expansion and obtain a near-scale invariant gauge field power spectrum we require $p \ge p_c$. The particular case $p=p_c$, which we refer to as the critical case, is very interesting. As we shall see, this is the case in which $R=0$ to first order in slow-roll parameter while the gauge field excitations become scale-invariant. As a result the gauge field plays the role of an isocurvature field during inflation. In this limit, there is no anisotropy generated during inflation and all anisotropies, as in \cite{Yokoyama:2008xw}, are generated at the surface of end of inflation via the inhomogeneous end of inflation effect.

We have chosen the convention that at the time of end of inflation, $\eta= \eta_f$, the conformal factor becomes unity and one restores the standard isotropic FRW universe after inflation. This means that the ansatz Eq. (\ref{f-a-scale}) or Eq. (\ref{f-eq}) is valid only during inflation, $\eta \leq  \eta_f$ and  $\phi \ge \phi_f$.
One has to provide a dynamical mechanism to stabilize the gauge kinetic coupling to a finite perturbative value at the end of inflation and during the reheating period.  We follow the phenomenological approach that
this can be achieved, at least in principle, without affecting the predictions of inflation on super-horizon scales.


As shown in \cite{Watanabe:2009ct} during the attractor regime
$R$ is given by
\ba
\label{R-app}
R = \frac{c-1}{2c}\epsilon = \frac{1}{2}I\epsilon \, ,
\ea
where $\epsilon\equiv -\dot H/H^2$ is the slow-roll parameter and we have defined the anisotropy parameter $I$ via
\ba
I\equiv\frac{c-1}{c} = \frac{p}{p_c}-1 \, .
\ea
 Combined with the definition of $R$ in Eq. (\ref{R-def}) we obtain
\ba
\label{cons-eq}
\dot A^2 f^2 e^{-2 \alpha} = I \, \epsilon V \, .
\ea
As we shall see below, this equation will be the key equation to find $\delta N$ in terms of $\delta \phi$ and $\delta \dot A$.

During the attractor phase the inflaton evolution is given by  \cite{Watanabe:2009ct}
\ba
\label{klinGordon}
M_P^{-2}\frac{d \phi}{d \alpha} \simeq -\frac{V_\phi}{ V} + \frac{c-1}{c}\frac{V_\phi}{ V} \, .
\ea
Interestingly, this equation means that the back-reactions of the gauge field on the inflation field change the effective mass of the inflaton field as given by the second term above.


\subsection{Solving for N}

Having specified the form of $f(\phi)$ we are ready to solve $N$ in terms of background fields dynamics. First, let us calculate $N$ as a function of the scalar field $\phi$. As described before, during the attractor phase the gauge field's back-reaction on inflaton dynamics is not negligible. The evolution of $\phi$ as a function of number of e-folds is given in
Eq. (\ref{klinGordon}) which for our potential yields
\ba
\label{phi-eq1}
\frac{d \phi}{d \alpha} &&= \left( I -1 \right) \frac{ M_{P}^{2}V_{, \phi}}{V}  \nonumber\\
&&= \left( I -1 \right) \frac{2 \phi }{p_c} \, .
\ea
Note that the effects of gauge field back-reaction on inflaton dynamics is captured by
the factor $I$ in Eq. (\ref{phi-eq1}). If we set $I=0$, then the gauge field has no effect on inflaton dynamics and one recovers the standard formula for $N(\phi)$ as in conventional hybrid inflation scenarios. Note that this is the critical limit in which the gauge field plays the role of an iso-curvature field during inflation. In other words, in the critical case we have turned on the conformal factor $f(\phi)$ such that  the gauge field fluctuations barely survive the background expansion and becomes exactly scale-invariant.

One can easily solve Eq. (\ref{phi-eq1}) to obtain
\ba
\label{N-phi-hybrid}
N(\phi) \simeq \frac{p_c}{2 \left(I- 1  \right) } \ln \left( \frac{\phi}{\phi_{f}} \right)  \, .
\ea
Note that we have replaced $\alpha = N$ and used the convention that $N$ is counted from the time of end of inflation $N_f=0$ such that  at the start of inflation where $\phi=\phi_i$, we have $N=N_i \simeq -60$. Eq. (\ref{N-phi-hybrid}) is one of our starting key
formula to find the final $\delta N$ expression.

Now we calculate $N$ as a function of the background gauge field dynamics. The key information is that during the attractor phase $R$ reaches a constant value as given by
Eq. (\ref{R-app}).  Plugging Eq. (\ref{R-app}) for definition of $R$ in Eq. (\ref{R-def}) and noting that
$f \simeq (a/a_f)^{-2} = e^{-2N}$ we obtain
\ba
\label{A-N}
A =  \sqrt{\frac{2R}{3}} M_P \left( e^{3 N} -1 \right) + A_{f} \, ,
\ea
in which $ A_f$ denotes the value of the gauge field at the end of inflation. From Eq. (\ref{A-N}) we see that the gauge field is completely  negligible during the most of the period of inflation  in which  $N<0$ and grows exponentially towards the end of inflation when it approaches  its final value $ A_f $.  Note that in the critical case in which $R \rightarrow 0$, we have $A \simeq A_f$ as expected for an iso-curvature field.

Alternatively, one can invert Eq. (\ref{A-N}) to find $N$ as a function of $A$
\ba
\label{N-A-hybrid}
N=  \frac{1}{3} \ln  \left( \sqrt{\frac{3}{2 R} } \left( \frac{ A -  A_{f} }{M_{P}} \right) + 1 \right) \, .
\ea
This is our second key formula to express the final formula of $\delta N$ in terms of the background fields perturbations.

Our last task in finding $\delta N$ is to take into account the contributions from the surface of the end of inflation  in $\delta N$. Here we follow closely the method developed in multi-brid inflation \cite{Sasaki:2008uc, Naruko:2008sq}. We parameterize the surface of end of inflation in Eq. (\ref{transition}) via \cite {Emami:2011yi}
 \ba
 \label{gamma}
 \phi_{f} = \phi_c \cos \gamma \quad , \quad
 A_{f} =  \frac{g\,  \phi_c}{\mathbf{e}} \sin \gamma \, .
\ea
We note that $\gamma$ should be treated as an independent variables so when we calculate
$\delta N$, we have to take into account the perturbations in $\gamma(\phi, A)$ which represents the contribution of the surface of the end of inflation in $\delta N$.

\subsection{ Calculating $\delta N$}
\label{deltaN }

Our goal is to calculate $\delta N$ up to second order in  $\delta \phi$ and $\delta A$ perturbations. We relegate the details of the analysis to
Appendix \ref{deltaN-app}. However, before using $\delta N$ formalism in this setup, one has to verify the validity of the gradient expansion and the separate Universe approach as studied in \cite{ Lyth:2004gb,  Naruko:2012fe, Naruko:2012um, Sugiyama:2012tj}. We demonstrate the validity of the separate Universe approach for our setup to second order in perturbation theory  in Appendix \ref{separate-universe}. This will allow us to use $\delta N$ formalism to
calculate the power spectrum and the bispectrum.

Calculating $\delta N( \phi, A)$ up to second order we have
\begin{align}
\label{delta N total}
\delta N &= N_{{\phi}} \delta {\phi} + N_{\dot{A}} \delta \dot{A} + N_{{A}} \delta {A}+ \frac{N_{{\phi} {\phi}}}{2} {\phi}^2  +  \frac{N_{\dot{A} \dot{A}}}{2} \delta \dot{A}^2 + \frac{N_{{A}{A}}}{2} \delta {A}^2\nonumber\\
&~~~+ N_{{\phi} \dot{A}} \delta {\phi} \delta \dot{A_x} +  N_{{\phi} {A}} \delta {\phi} \delta {A_x} + N_{{A} \dot{A}} \delta \dot{A} \delta {A} \, ,
\end{align}
in which to leading order in $R \ll 1$, for linear terms we have
\begin{align}
\label{N-phi}
 N_{{\phi}} & =   \frac{p_{c}}{2(I-1)} \frac{1}{{\phi}}   + \frac{\e M_{P}p_{c}}{ 6 \phi_c \, g \, (I-1)}\frac{\tan{\gamma}}{\cos{\gamma}} \frac{f_{\phi}}{f} \sqrt{ 6 R} \\
 \label{N-dotA}
 N_{\dot {A}} & = \left(-\frac{2NI}{(I-1)} +  \frac{ \e \, p_{c}}{6 g (I-1)}\frac{\tan{\gamma}}{\cos{\gamma}} \frac{M_{P}}{\phi_{c}} \sqrt{ 6 R} \right) \left( \frac{1}{\dot {A}}\right) \\
 \label{N-A}
  N_{{A}} & = \frac{\e \, p_{c}}{2 g \phi_c (I-1)}\frac{\tan{\gamma}}{\cos{\gamma}} \, .
  \end{align}
We note that in above expressions, the first term in $N_\phi$ represents the usual contribution in curvature perturbations from the inflaton field except with the small correction from the gauge field back-reaction given by the factor $I$ in the denominator.  The second term in
$N_\phi$, proportional to $\e$, represents the contribution in isotropic power spectrum from the inhomogeneities generated at the surface of end of inflation. On the other hand, the first term in $N_{\dot A}$, containing $ N I$, represents the anisotropies generated actively during inflation from $\delta \dot A_i$ fluctuations. This is the crucial contribution first derived in the analysis of  \cite{Abolhasani:2013zya} in the $\delta N$ formalism. Furthermore, the second term in $N_{\dot A}$ containing $\e$, represents the anisotropies generated purely at the surface of end of inflation again from the $\delta \dot A_i$ fluctuations. This is obtained  for the first time in this work in the context of $\delta N$ formalism. As we shall see, this is the  term which  captures the anisotropies generated from the surface of end of inflation as envisaged by Yokoyama and Soda in \cite{Yokoyama:2008xw}.  Finally, the term $N_A$ represents the anisotropies generated purely from the surface of end of inflation from  $\delta A_i$ fluctuations. This is the term which was included in the analysis of \cite{Emami:2011yi}. As we shall see below, the contribution from $N_A$ term is exponentially suppressed compared to the contribution from $N_\phi$ and $N_{\dot A}$, consistent with the  analysis of \cite{Emami:2011yi}. However, in the analysis of \cite{Emami:2011yi} the crucial contribution of $N_{\dot A}$, Eq. (\ref{N-dotA}),  was not included which has led to the conclusion different than the result obtained in \cite{Yokoyama:2008xw}.

Correspondingly, for the non-linear terms we get
  \ba
  N_{{\phi} {\phi}} & =&  - \frac{p_{c}}{2(I-1)} \frac{1}{{\phi}^2}  - \frac{2NI}{(I-1)} \left( \left(\frac{f_{\phi}}{f}\right)^2+ \frac{f_{,\phi \phi}}{f}\right) + \frac{\e^2 M_{P}^2 p_{c}}{6 g^2 \phi_c^2 (I-1)}\frac{(\sin^2{\gamma}+1)}{\cos^4{\gamma}} \left(\frac{f_{\phi}}{f} \right)^2 I \epsilon \nonumber \\
  &&~~  -\frac{2 \e Ip_{c}}{3 g (I-1)^2}\frac{\tan{\gamma}}{\cos^2{\gamma}} \frac{ M_{P}f_{\phi}^2}{\phi_c f^2} \sqrt{ 6 R}+
\frac{\e p_{c}}{2g (I-1)}\frac{\tan{\gamma}}{\cos{\gamma}} \frac{ M_{P}f_{\phi \phi}}{f \phi_c} \sqrt{\frac{ 2R }{3}} \\
 N_{{A} {A}} & =&  \frac{\e^2 p_{c}}{2 g^2 \phi_c^2(I-1)}\frac{(\sin^2{\gamma}+1)}{\cos^4{\gamma}} \\
  N_{\dot {A}\dot {A}} & = &\left(-\frac{2NI}{(I-1)} + \frac{ \e^2 \, R\,  p_{c}}{3 g^2 (I-1)}\frac{(\sin^2{\gamma}+1)}{\cos^4{\gamma}} \frac{M_{P}^2}{\phi_{c}^2} -\frac{2 \e I p_{c}}{3g (I-1)^2}\frac{\tan{\gamma}}{\cos{\gamma}} \frac{M_{P}}{\phi_{c}}
  \sqrt{6 R} \right) \left( \frac{1}{\dot {A}}\right)^2
  \ea
  \ba
  N_{{A} {\phi}} & = & \frac{\e^2 p_{c}}{6 g^2 (I-1)}\frac{(\sin^2{\gamma}+1)}{\cos^4{\gamma}}\frac{M_{P}f_{\phi}}{\phi_c f} \sqrt{ 6 R} - \frac{ \e I p_{c}}{g (I-1)^2}\frac{\tan{\gamma}}{\cos{\gamma}} \frac{f_{\phi}}{\phi_c f} \\
   N_{{A} \dot{A}} & =& \left(\frac{\e^2 p_{c}}{6 g^2 (I-1)}\frac{(\sin^2{\gamma}+1)}{\cos^4{\gamma}}\frac{M_{P}}{\phi_{c}} \sqrt{6 R} - \frac{ \e I p_{c}}{g (I-1)^2}\frac{\tan{\gamma}}{\cos{\gamma}}\right)  \left( \frac{1}{\dot {A}}\right) \\
  N_{\dot{A} {\phi}} & =& - \frac{I p_{c}}{(I-1)^2} \frac{1}{{\phi}}  - \frac{4NI}{(I-1)} \frac{f_{\phi}}{\phi_c f} + \frac{\e^2 M_{P}^2 p_{c}}{6 g^2 (I-1)\phi_{c}^2}\frac{(\sin^2{\gamma}+1)}{\cos^4{\gamma}} \frac{f_{\phi}}{f} I \epsilon \nonumber \\
\label{N-dotA-phi}
 & &~~  + \frac{ \e p_{c}}{6 g (I-1)}\frac{\tan{\gamma}}{\cos{\gamma}} \frac{ M_{P}f_{\phi}}{\phi_c f} \sqrt{6 R} \, ( 1- 4 I)  \, .
\ea

For the future reference, we note that if we set $I=0$ in the above expressions, we obtain
the results in the critical limit $p=p_c$ in which the gauge field does not back-react on the inflaton dynamics and its contributions into anisotropies are generated purely from the surface of end of inflation controlled by the gauge coupling $\e$.

\subsection{Seed Quantum Fluctuations}

Having calculated $\delta N$ to second order in fields perturbations we are ready to calculate the power spectrum and the bispectrum. Before doing that, we need some information about the statistical properties of $\delta \phi$ and $\delta A$ fluctuations necessary for
$\delta N$ analysis. As usual, we take $\delta \phi$ to be a near scale-invariant perturbations with amplitude $H/2\pi$ at the time of horizon crossing such that
\ba
\label{dis phi}
\left<\delta  \phi_{\bfk} \delta \phi_{\bfk'}\right> \equiv (2\pi)^{3} P_{\delta  \phi}(k)~\delta^3(\bfk+\bfk') ~~,~~ {\cal P}_{\delta  \phi}\equiv \frac{k^{3}}{2 \pi^{2}}P_{\delta  \phi}(k) = \left( \frac{H_*}{2 \pi} \right)^2 \, .
\ea
Note that  this is calculated at the time when the mode of interest leaves the horizon, denoted by  $*$,  when $k=a_* H_*$.

To find the power spectrum of gauge field fluctuations  we solve the mode function for $\delta A_i$ with the initial Bunch-Davies vacuum when the mode of interest is deep inside the horizon.   The canonically normalized gauge field quantum fluctuations are decomposed via
\cite{Bartolo:2012sd}
\ba
\label{A-V-eq}
\delta A_i =   \sum_{\lambda = \pm} \int \frac{d^3k}{(2\pi)^{3/2}}e^{i \bfk.\mathbf{x}}\vec{\epsilon}_{\lambda}(\bfk) \frac{\widehat{V}_i}{f}  \, ,
\ea
where
\ba
\widehat{V} = a_{\lambda}(\bfk)V_{\lambda}(k) + a_{\lambda}^{\dagger}(-\bfk)V_{\lambda}^{*}(k) \, .
\ea
As usual  $a_{\lambda}(\bfk)$ and $a_{\lambda}^{\dagger}(\bfk)$
represent the annihilation and the creation operators. Furthermore,  $\epsilon_\lambda$ represents the circular polarization for $\lambda =\pm$,
satisfying  $\vec k \,  . \,   \vec \epsilon_\pm(\bfk) =0 \, , \,  \bfk \, \times \,  \vec \epsilon_\pm(\bfk) =
\mp i k \, \vec \epsilon_\pm(\bfk)  \, , \vec \epsilon_\lambda (-\bfk) = \vec \epsilon_\lambda\, (\bfk)^*$, normalized such that
$\vec \epsilon_\lambda(\bfk)  \, . \,  \vec \epsilon_{\lambda'}(\bfk)^*= \delta_{\lambda \lambda'}$ and
\ba
\sum_{\lambda}\epsilon_{\lambda,i}({\bfk})\epsilon^{*}_{\lambda,j}({\bfk}) = \delta_{ij} - \hat{k}_{i}\hat{k}_{j} \, .
\ea
Finally,  the mode functions $V_\lambda(\bfk)$ satisfies the evolution equation
\begin{align}
\label{canonical gauge field equation}
V_{\lambda}(k)''+ \left( k^2 - \frac{f''}{f} \right)V_{\lambda}(k)=0 \, ,
\end{align}
in which  a prime denotes the derivative with respect to conformal time $d \eta= dt/a(t)$.
With  $f$ given in  Eq. (\ref{f-eq}) the normalized gauge field mode function has the same form as that of a   a massless scalar field in dS background
\begin{align}
\label{canonical gauge field}
V_{\lambda}(k) \simeq \frac{1+ i k \eta}{\sqrt{2}k^{3/2}\eta} e^{-ik\eta} \, .
\end{align}
As a result, on super-horizon scales we obtain \cite{ Abolhasani:2013zya}
\ba
\label{mode-deltaA}
\frac{\delta \vec{\dot{A}}}{\dot{A}} = \frac{1}{M_P} \sqrt{\frac{3}{2 R}}
\frac{H}{ \sqrt{2 k^3}}
\sum_{\lambda}\vec{\epsilon}_{\lambda}     \quad \quad (k > a H)
\ea
 Eq. (\ref{mode-deltaA}) shows that  $ \delta \dot A/\dot A $ is scale-invariant on super-horizon scales which is crucial for our analysis below. This is a consequence of turning on the
 conformal factor $f(\phi)$ with the specific form given in Eq. (\ref{f-eq}).

One crucial point to mention is that although $V_\lambda(k)$ has the profile of a massless scalar field fluctuations as given in Eq. (\ref{canonical gauge field}), but the profile of gauge field
$\delta A$ is related to $V_\lambda(k)$ by additional factor of $f(\phi)^{-1}$ as shown in
Eq. (\ref{A-V-eq}). As a result, the power spectrum of $\delta A$, $\calP_{\delta A}$, is hugely suppressed compared to $\calP_{\delta \phi}$. More specifically
\ba
\label{P-A-less}
\calP_{\delta A_*} = \frac{1}{f(\phi_*)^2} \calP_{\delta \phi} \simeq e^{4 N_*}  \calP_{\delta \phi_*} \, .
\ea
Noting that $N_* \simeq -60$, we conclude that $\calP_{\delta A_*} \ll \calP_{\delta \phi}$.
This was the key argument in  \cite{Emami:2011yi} as why the  method employed in \cite{Yokoyama:2008xw} does not work. As we discussed in previous sub-section, in the paragraph after Eq. (\ref{N-A}),  in order to reproduce the results of \cite{Yokoyama:2008xw} correctly, we have to take into account
the contribution of $\delta \dot A/\dot A$, represented by $N_{\dot A}$ in Eq. (\ref{N-dotA}),
which is scale-invariant as given in Eq. (\ref{mode-deltaA}) following the method developed  in \cite{ Abolhasani:2013zya}. This was the crucial missing piece in the analysis of \cite{Emami:2011yi}.


\section{Power Spectrum}
\label{power-spec}

Having calculated $\delta N$ to second order in Eq. (\ref{delta N total}) we can calculate the power spectrum and the bispectrum using the standard formula
\ba
\calR(\mathbf{x},t) = \delta N( \phi, A) \, .
\ea
The power spectrum analysis are presented here while the bispectrum analysis are given in next Section.

The quantities $N_\phi, N_A$ and $N_{\dot A}$ are given in Eqs. (\ref{N-A}), (\ref{N-dotA})
and (\ref{N-A}). As a result
\ba
\calP_\calR = N_\phi^2 \calP_{\delta \phi} + \dot A^2 N_{\dot A}^2 \calP_{\delta \dot A/\dot A}
+ N_{A}^2 \calP_{\delta A}
\ea
As argued in Eq. (\ref{P-A-less}), one can safely neglect the contribution of $\calP_{\delta A}$ in the above formula since it is exponentially suppressed compared to $\calP_{\delta \phi}$.
This is reminiscent of the argument used in \cite{Emami:2011yi}. However, as will show, the source of anisotropy is actually from $\calP_{\delta \dot A/\dot A}$. Neglecting the contribution of $\calP_{\delta A}$ we obtain
\ba
\calP_\calR \simeq  \calP_{\calR }^{(0)}  + \dot A^2 N_{\dot A}^2 \calP_{\delta \dot A/\dot A} \, ,
\ea
in which the isotropic power spectrum is calculated to be
\ba
\label{power}
\calP_{\calR }^{(0)} = N_{\phi_*}^2 \calP_{\delta \phi_*} = \left( \frac{p_c H_*}{4 \pi \phi_*}
\right)^2 \left[ 1+ \frac{\e p_c M_P}{g \phi_c} \frac{\tan \gamma}{\cos \gamma} \sqrt{\frac{2 R}{3}} \right]^2   \, ,
\ea
in which an asterisk represents the time of horizon crossings, $k= a_* H_*$, and
the relation $f_\phi/f \simeq p_c/\phi$ has been used. Furthermore, it is assumed that
$I \ll 1 $ which follows from the observational constraint as we shall see below,
see also  \cite{ Watanabe:2010fh, Abolhasani:2013zya}.

Now we calculate the anisotropy generated during and at the surface of end of inflation coming from $\calP_{\delta \dot A/\dot A}$. First, from Eq. (\ref{mode-deltaA}),  we note that
\ba
\calP_{\delta \dot A/\dot A} = \frac{k^3}{2 \pi^2 }
\left\langle  \frac{\delta \dot A_x(k_1)}{A_x}
\frac{\delta \dot A_x(k_2)}{A_x} \right\rangle = \frac{3 H_*^2 \sin^2 \theta }{8 \pi^2 M_P^2 R} \, ,
\ea
in which $\cos \theta = \hat {\bf n}. \hat{\bf k}$ represents the angle between the preferred direction (in our notation the $x$-direction) and the wave-number.  Using the expression for
$N_{\dot A}$ given in Eq. (\ref{N-dotA}), the anisotropy  in power spectrum becomes
\ba
\Delta \calP_\calR \equiv \dot A^2 N_{\dot A}^2 \calP_{\delta \dot A/\dot A}  =
 \left(-2NI  +  \frac{ \e \, p_{c}}{6 g }\frac{\tan{\gamma}}{\cos{\gamma}} \frac{M_{P}}{\phi_{c}} \sqrt{ 6 R} \right)^2  \frac{3 H_*^2 \sin^2 \theta }{8 \pi^2 M_P^2 R} \, .
\ea
On the other hand, using the relation $R= I \epsilon/2 $ and
\ba
\label{epsilon-eq}
\epsilon \simeq \frac{2 \phi_*^2}{p_c^2 M_P^2}
\ea
and assuming $\phi_* \simeq \phi_f$ which is a good approximation in hybrid inflation,
the fractional change in power spectrum due to anisotropies from the gauge fields is
\ba
\frac{\Delta \calP_\calR}{\calP_{\calR }^{(0)}} =  \left( \frac{\sqrt{24 I N^2} - \frac{\e}{g} \tan \gamma}{1+ \frac{\e p_c M_P}{g \phi_c} \frac{\tan \gamma}{\cos \gamma} \sqrt{\frac{2}{3} R}} \right)^2  \sin^2 \theta \, .
\ea

Now, in the limit in which $R= I \epsilon/2 \ll1 $ (assuming the pre-factor in denominator above does not diverge for typical value of $\e$ and $\gamma$) we obtain
\ba
\label{g-eq}
g_* \simeq  -\left( \sqrt{24 I N^2} - \frac{\e}{g} \tan \gamma \right)^2 \, .
\ea
This is one of the main result of this paper.

Eq. (\ref{g-eq}) is very interesting. The first term in the bracket captures the effects of anisotropy generated actively during inflation \cite{ Watanabe:2010fh, Bartolo:2012sd,
Abolhasani:2013zya, Emami:2013bk}.
As is well-known the anisotropy generated during inflation is given by
\ba
\label{g-IR}
g_*|_{IR} = -24 I N^2
\ea
in which the subscript IR indicates the accumulative contributions of the IR modes which leave the horizon at  $N$ e-folds towards the end of inflation. As mentioned in \cite{Bartolo:2012sd} the IR contribution can become
too large so in order not to produce too much anisotropy from IR accumulations one requires
that $N$ is not too large. With $N=60$  to solve the flatness and the horizon problem, and $| g_* | \lesssim 10^{-2}$ in order to satisfy the observational bounds, one requires $I \lesssim 10^{-7}$.

The second term in Eq. (\ref{g-eq}) represents the anisotropy generated purely from the surface of end of inflation. This is the contribution advocated originally in
\cite{Yokoyama:2008xw}. Denoting this contribution by $g_*|_{\e}$ we have
\ba
\label{ge}
g_*|_{\e} = - \frac{\e^2}{g^2} \tan^2 \gamma \, .
\ea
Note that this term exists even we set $I=0$. This is the critical limit in which the gauge field becomes an iso-curvature field which does not contribute into power spectrum
during inflation but can generate additional anisotropic curvature perturbations from the
inhomogeneities generated at the surface of end of inflation a la Lyth \cite{Lyth:2005qk}.
We can also compare our expression for $g_*|_{\e}$ with the value for $g_*$ obtained in
\cite{Yokoyama:2008xw}, see also \cite{ Emami:2011yi} and \cite{Lyth:2012vn}.
Interestingly, we find that in the physical limit in which $| g_*| \ll 1$, our value for $g_*|_{\e}$ agrees with $g_*$ obtained in  \cite{Yokoyama:2008xw}.
This may seem somewhat surprising, since in the analysis of \cite{Yokoyama:2008xw} the exponential evolution of the gauge field during inflation and the the effects of $N_{\dot A}$ in $\delta N$ expansion is not taken into account. However, the analysis of \cite{Yokoyama:2008xw} relies crucially on the idea of \cite{Lyth:2012vn} in which the inhomogeneities at the surface of end of inflation are generated from a scale-invariant perturbation, which in our
case it is $\delta \dot A/\dot A$. The fact that the fluctuations $\delta \dot A/\dot A$ are massless and scale-invariant guarantees that the method proposed in \cite{Lyth:2005qk} still applies in the analysis of \cite{Yokoyama:2008xw}. This is the reason why the results obtained in \cite{Yokoyama:2008xw} are correct as a matter of principle independent of
their somewhat ad-hoc $\delta N$ analysis.

One curious conclusion from  Eq. (\ref{g-eq}) is that  one can choose the parameter space such that  the anisotropies generated actively during inflation are cancelled from the anisotropies generated from the surface of end of inflation, yielding $g_*=0$. For this to happen one requires
\ba
\label{tuning}
\tan \gamma = \frac{g}{\e} \sqrt{24 I N^2} \, .
\ea
 For a given value of number of e-foldings $N$, this determines how one should approach the surface of end of inflation in order to obtain $g_* =0$. Of course, in this limit, one should take into account the sub-leading terms containing factors of $I$ in $1-I $ expressions and so on
 in $\delta N$ expression in previous sub-section. Nonetheless, it may be possible to achieve $g_*=0$ to higher order in $I$ too.

 One interesting implication of  Eq. (\ref{g-eq}) is that one does not need to impose the strong fine-tuning $I \lesssim 10^{-7}$ anymore. The smallness of $g_*$ is now controlled by the near cancellation of the two terms in  Eq. (\ref{g-eq}) instead of requiring the first term, containing $\sqrt {I N^2}$, itself to be small.

Now let us look at the spectral tilt, $n_s$.  Similar to \cite{Ohashi:2013qba} we have
\ba
\label{ns}
n_s -1= \frac{d \ln \calP_\calR}{d \ln k} = ( n_s^{(0)} -1  ) + \frac{\cos^2 \theta }{1+ g_* \cos^2 \theta } \frac{d g_*}{d \ln k} \, ,
\ea
in which $n_s^{(0)}$ represents the spectral index obtained from the isotropic power spectrum. In conventional models of hybrid inflation in which the waterfall is a real field and is not gauged under the $U(1)$ field $n_s^{(0)}>1$ so the power spectrum is blue-tilted. As a result, the simple models of hybrid inflation are ruled out from the PLANCK data which requires $n_s <1$. On the other hand,  with $g_*$ given in Eq. (\ref{g-eq}), the change in spectral index induced by anisotropies, $\Delta n_s$, in the limit $| g_*| \ll 1$ is obtained to be
\ba
\label{delta-ns}
\Delta n_s \simeq - \sqrt{-96 I g_*}  \, \langle \cos^2 \theta \rangle =- \frac{ \sqrt{-96 I g_*}}{3} \, ,
\ea
in which we have taken the average value $\langle \cos^2 \theta \rangle = 1/3$.

Interestingly, anisotropies, once averaged over the whole sky, generate a red-tilted power spectrum. This may help to generate a red-tilted power spectrum for hybrid inflation.  As an example, suppose we take $g_* \sim -10^{-2}$ and $I \sim 10^{-2}$, then we obtain $\Delta n_s \sim - 0.03$
which can bring the value of $n_s$ in models of  gauged hybrid inflation within the desired range of PLANCK data. Having this said, this value of $I$ may seem too large compared to results obtained in \cite{ Watanabe:2010fh, Bartolo:2012sd,
Abolhasani:2013zya, Emami:2013bk}. However, in those works anisotropies are generated during inflation so $g_*$ is given in Eq. (\ref{g-IR}), yielding $I \sim 10^{-7}$. However, in our model, as we mentioned before, anisotropies generated at the surface of end of inflation can nearly cancel the anisotropies generated during inflation so one can saturate the observational bound on $g_*$ without requiring very small value of $I$. Therefore, taking
a value of $I$ much larger than the value required in \cite{ Watanabe:2010fh, Bartolo:2012sd,  Abolhasani:2013zya, Emami:2013bk} is easy in our model.

Before ending this section we comment that we have neglected the contributions of the waterfall quantum fluctuations and the longitudinal excitation in curvature perturbations. The reason is that the waterfall is very massive during inflation and its power spectrum is highly blue-tilted. As studied in great details in  \cite{Abolhasani:2010kr, Abolhasani:2011yp,      Fonseca:2010nk, Gong:2010zf, Lyth:2010ch, Lyth:2012yp} the contributions of the waterfall on large scale (i.e. CMB scale) curvature perturbations are exponentially suppressed. Similarly, the longitudinal mode is very massive and its excitations are highly suppressed compared to the transverse modes. As demonstrated in \cite{Emami:2013bk} in a similar model,
the contributions of the longitudinal excitation in curvature perturbations are exponentially suppressed compared to the transverse mode.

\section{BiSpectrum}
\label{bi-spec}

In this Section we calculate the Bispectrum of our model. For similar analysis in anisotropic inflation background of \cite{Watanabe:2009ct} see \cite{Bartolo:2012sd, Shiraishi:2013vja}
As demonstrated in \cite{Abolhasani:2013zya}
the main contribution in bispectrum comes from the $N_{\dot A \dot A}$ terms. The contributions from terms containing $N_{\phi \phi}$ are slow-roll suppressed as demonstrated by Maldacena \cite{Maldacena:2002vr}. Furthermore, similar to power spectrum case, terms containing $N_{AA}, N_{A \dot A}$  and $N_{A \phi}$ are suppressed as argued in Eq. (\ref{P-A-less}). In addition, one can check that the contribution from the term
$N_{\dot A \phi}$ is suppressed by a factor of $1/N$  compared to the contribution from
the  $N_{\dot A \dot A}$ term.  As a result, in our analysis below we consider the contributions from $N_{\dot A \dot A}$ term.

Calculating the three point function, and taking $I \ll 1$,   we obtain
\ba
\label{leading three0}
\left \langle \calR (\bfk_{1}) \calR (\bfk_{2}) \calR (\bfk_{3})   \right \rangle  &\simeq& \frac{1}{2}
N_{, \dot{A}\dot{A}}( k_{1}) N_{, \dot{A}}( k_{2}) N_{, \dot{A}} ( k_{3}) \int \frac{d^3\bfp}{(2\pi)^3} \left \langle \delta \dot{A}_{x} (\bfk_{1})  \delta \dot{A}_{x} (\bfk_{2})
\delta \dot{A}_{i} (\bfp) \delta \dot{A}_{i} (\bfk_{3} - \bfp)  \right \rangle + 2 \mathrm{perm.} \nonumber\\
&&=   \left(N_{k_3} I - \frac{\e^2 \,  R \, p_{c}}{6 g^2}\frac{(\sin^2{\gamma}+1)}{\cos^4{\gamma}} \frac{M_{P}^2}{\phi_{c}^2} - \frac{ \e\,  I p_{c}}{3 g }\frac{\tan{\gamma}}{\cos{\gamma}} \frac{M_{P}}{\phi_{c}}\sqrt{3I \epsilon} \right)\times \nonumber \\
&& \times \left(2N_{k_1} I - \frac{\e \, p_{c}}{6 g }\frac{\tan{\gamma}}{\cos{\gamma}} \frac{M_{P}}{\phi_{c}} \sqrt{3I \epsilon} \right)\Bigg{(}2N_{k_2} I - \frac{\e\,  p_{c}}{6 g }  \frac{\tan{\gamma}}{\cos{\gamma}} \frac{M_{P}}{\phi_{c}} \sqrt{3I \epsilon} \Bigg{)} \nonumber\\
 &&\times \int \frac{d^3\bfp}{(2\pi)^3}  \left\langle \frac{\delta \dot{A}_{x} (\bfk_{1})}{\dot{A}}  \frac{\delta \dot{A}_{x} (\bfk_{2})}{\dot{A}}\frac{\delta \dot{A}_{i} (\bfp)}{\dot{A}}
\frac{\delta \dot{A}_{i} (\bfk_{3} -\bfp)}{\dot{A}} \right\rangle  + 2 \mathrm{perm.} \, ,
\ea
in which $N_{k_i}$ represents the time when the mode $\bfk_i$ leaves the horizon. Calculating the convolution integrals and defining the bispectrum $B_{\calR} (\bfk_1, \bfk_2, \bfk_3)$
via
\ba
\label{bi- def}
\langle \calR (\bfk_{1}) \calR (\bfk_{2}) \calR (\bfk_{3}) \rangle &\equiv& \left( 2 \pi \right)^3 \delta^3 \left( \bfk_{1} +  \bfk_{2} +  \bfk_{3}\right) B_{\calR}(\bfk_{1}, \bfk_{2}, \bfk_{3})
\ea
 we get
\ba
\label{leading three}
B_{\calR}(\bfk_{1}, \bfk_{2}, \bfk_{3})   &\simeq& \frac{18 \epsilon^2}{R^2}
 \left(2N_{k_1} I - \frac{\e \, p_{c}}{6 g }\frac{\tan{\gamma}}{\cos{\gamma}} \frac{M_{P}}{\phi_{c}} \sqrt{6 R} \right)\Bigg{(}2N_{k_2} I - \frac{\e\,  p_{c}}{6 g }  \frac{\tan{\gamma}}{\cos{\gamma}} \frac{M_{P}}{\phi_{c}} \sqrt{6 R} \Bigg{)}
\times  \\
&& \left(N_{k_3} I - \frac{\e^2 \,  R \, p_{c}}{6 g^2}\frac{(\sin^2{\gamma}+1)}{\cos^4{\gamma}} \frac{M_{P}^2}{\phi_{c}^2} - \frac{ \e\,  I p_{c}}{3 g }\frac{\tan{\gamma}}{\cos{\gamma}} \frac{M_{P}}{\phi_{c}}\sqrt{6R} \right)  
\Big( C(\bfk_{1}, \bfk_{2})P_0(k_{1})P_0(k_{2}) + 2 \mathrm{perm.} \Big) \, , \nonumber\\
\ea
in which $P_0(k)= (2 \pi^2/k^3) \calP_{\calR}^{(0)}(k)$ and
the momentum shape function $C(\bfk_1, \bfk_2) $ is defined via  \cite{ Bartolo:2012sd,  Abolhasani:2013zya}
\ba
C(\bfk_{1}, \bfk_{2})\equiv\bigg{(}1 -   (\widehat \bfk_1.\widehat {\bf{n}} )^2  -   (\widehat \bfk_2.\widehat {\bf n})^2 +
(\widehat \bfk_1.\widehat {\bf n} ) \,  (\widehat \bfk_2.\widehat {\bf n}) \,  (\widehat \bfk_1.\widehat \bfk_2)  \bigg{)} \, .
\ea
As in the power spectrum case, the bispectrum is generated both during inflation and at the end of inflation. The bispectrum generated during inflation in Eq. (\ref{leading three}) are represented
by $N_{k_i} I $ factor while the bispectrum generated from inhomogeneous end of inflation effect are controlled by the gauge coupling $\e$.

Now let us look at the bispectrum generated purely during inflation by setting $\e=0$ in Eq. (\ref{leading three}). In this limit, we obtain
\ba
\label{B-active}
B_{\calR}(\bfk_{1}, \bfk_{2}, \bfk_{3}) |_{IR} = 288 I\, N_{k_1} N_{k_2} N_{k_3}
\Big( C(\bfk_{1}, \bfk_{2})P_0(k_{1})P_0(k_{2}) + 2 \mathrm{perm.} \Big)\, \quad \quad
(\e=0) \, .
\ea
This is en exact agreement with the result obtained in  \cite{Bartolo:2012sd, Abolhasani:2013zya}.

Now let us look at the critical case in which the gauge field plays the role of an iso-curvature field, $I=0$, and the bispectrum is generated purely at the surface of end of inflation.  In this limit, we get
\ba
\label{B-cric}
B_{\calR}(\bfk_{1}, \bfk_{2}, \bfk_{3}) |_{\e} = -\frac{2\e^4 }{ p_c g^4} \frac{\sin^2 \gamma (1+ \sin^2 \gamma)}{\cos^4 \gamma} \Big( C(\bfk_{1}, \bfk_{2})P_0(k_{1})P_0(k_{2}) + 2 \mathrm{perm.} \Big)
\, , \quad \quad (p=p_c, I=0) \, .
\ea
Comparing Eqs. (\ref{B-active}) and (\ref{B-cric}), we find that the bispectrum generated at
the surface of end of inflation always has an opposite sign compared to  the bispectrum generated during inflation.

It is instructive to look at the non-Gaussian parameter $f_{NL}$ defined in the squeezed limit $k_1 \ll k_2 \simeq k_3$ via
\ba
f_{NL} (\bfk_1, \bfk_2, \bfk_3) = \lim_{k_1 \rightarrow 0} \frac{5}{12}
\frac{B_\zeta(\bfk_1, \bfk_2, \bfk_3)}{P_\zeta(k_1) P_\zeta(k_2)} \, .
\ea
If we further assume $N_{k_1} \simeq N_{k_2} \simeq N_{k_3}$ and noting that
$R = I \epsilon/2$ and using Eq. (\ref{epsilon-eq}), from  Eq. (\ref{leading three}) we obtain
\ba
\label{fNL-g*}
f_{NL} = -\frac{5 \epsilon g_*}{R} \left[ N I - \frac{\e I}{g}\sqrt{\frac{2 I}{3}} \tan \gamma
- \frac{\e^2 R}{3  p_c \epsilon g^2} \frac{1+ \sin^2 \gamma}{\cos^2 \gamma }
\right] C(\bfk_1, \bfk_2) \, .
\ea
This is an interesting result. Specially, consider the situation in which Eq. (\ref{tuning}) is satisfied so $g_*=0$. Then Eq. (\ref{fNL-g*}) also  indicates that $f_{NL}=0$. In other words, the anisotropies generated actively during inflation are canceled by the anisotropies generated from the surface of end of inflation both at the level of power spectrum and bispectrum.

In the limit $\e=0$ one obtains the known result that
\ba
\label{fnl squeezed}
f_{NL} (\bfk_1, \bfk_2)|_{IR}&=& 240 I N(k_{1}) N(k_{2})^2 C(\bfk_{1}, \bfk_{2})   \quad \quad  \quad
 ( k_{1} \ll k_{2} \simeq k_{3} )   \quad \quad (\e=0) \\
&\simeq& 10 N\,  |g_*| \, C(\bfk_1, \bfk_2)  \, . \nonumber
\ea
On the other hand, for the critical case with $I=0$ we obtain
\ba
\label{fNL-e0}
f_{NL}(\bfk_1, \bfk_2)|_\e =   -\frac{5\e^4}{3p_cg^4}  \frac{\sin^2 \gamma (1+ \sin^2 \gamma)}{\cos^4 \gamma}  C(\bfk_1, \bfk_2)
 &=&  -\frac{5 g_*^2}{3p_c} \frac{1+ \sin^2 \gamma}{\sin^2 \gamma }  C(\bfk_1, \bfk_2)
 \quad
 ( k_{1} \ll k_{2} \simeq k_{3} )
 \quad
  \quad (I=0)
\ea

To satisfy the observational bound from the PLANCK data, one should not produce too much non-Gaussianity. However, we note that the anisotropic non-Gaussianity given by the
shape function $C(\bfk_1, \bfk_2)$ is quite different than the known non-Gaussian shapes.
In the critical case in which $I=0$ and all anisotropies are generated at the surface of end of inflation, this is easy to satisfy. Indeed, Eq. (\ref{fNL-e0}) indicates that if $\gamma $ is not very close to zero, then $f_{NL}$ is quite negligible with small $g_*$. On the other hand, if  anisotropy is generated predominantly during inflation, then Eq. (\ref{fnl squeezed}) indicates that $f_{NL}$ is at the order of few which can be detectable.

Similar to the case of  power spectrum, Eqs. (\ref{B-cric}) and (\ref{fNL-e0}) have the interesting property that they doe not depend on the length of the duration of inflation. As a result, the IR  anisotropy problem associated with the model such as \cite{ Watanabe:2010fh, Bartolo:2012sd,  Abolhasani:2013zya, Emami:2013bk} in which
primordial anisotropies are generated during inflation does not show up  in models in which anisotropies are generated exclusively at the surface of end of inflation.

It is also constructive to compare our expression for $f_{NL}|_\e$ with $f_{NL}$ obtained
in \cite{Yokoyama:2008xw}. Taking the limit in which the direction-dependence in their formula collapses to our $C(\bfk_1, \bfk_2)$ their formula for $f_{NL}$ scales like
(in our notation)  $(\e^6/g^6) \tan^4 \gamma (1+ \tan^2 \gamma)$ which is different than our result obtained in Eq. (\ref{fNL-e0}).

\section{Summary and Discussions}
\label{summary}

In this work we have studied primordial quadrupole  asymmetry in models of gauged hybrid inflation. As we mentioned there are two mechanisms to generate primordial anisotropies from gauge fields: either actively from IR contributions during inflation or from the inhomogeneities
generated at the surface of end of inflation via the waterfall mechanism. The anisotropies generated during inflation suffers from the IR problem in which $g_*$ grows as $N^2$ which
leads to too much anisotropies to be compatible with the observation. Similarly, in this mechanism $f_{NL}$ scales like $N^3$ which is again problematic in the light of recent data
indicating no detectable non-Gaussianity. On the other hand, the mechanism of generating primordial anisotropy at the surface of end of inflation has the appealing feature that it does not suffer from the above mentioned IR problem. In this mechanism, the gauge field is an iso-curvature field which has negligible contribution in total energy density  so it does not affect the inflaton dynamics. However, it modulates the waterfall mechanism yielding to inhomogeneities at the surface of end of inflation. As a result, this generates primordial anisotropies purely at the end of inflation as pioneered by Yokoyama and Soda \cite{Yokoyama:2008xw}.

In this work we have employed a consistent $\delta N$ mechanism, similar to \cite{Abolhasani:2013zya}, to calculate the curvature perturbations up to second order. We have clearly identified the contributions in anisotropic power spectrum from the above two mechanisms as given in Eq. (\ref{g-eq}). The  two limiting cases
are given by Eqs. (\ref{g-IR}) and (\ref{ge}). Similarly, the different contributions in bispectrum are identified as given by Eq. (\ref{leading three}) with the two limiting cases
given in Eqs. (\ref{fnl squeezed}) and (\ref{fNL-e0}).

The combined effects of generating anisotropies during inflation and at the surface of end of inflation have some interesting observational implications. First, one can choose the parameter space such that these two source of anisotropies cancel each other's contributions so there is no net primordial anisotropies both at the level of power spectrum and bispectrum. Second, the anisotropic power spectrum are red-tilted. After averaging the anisotropic power spectrum over the sky, the total power spectrum can become red-tilted. This is an interesting observation which can save models of hybrid inflation which in the absence of gauge fields
predict a blue-tilted  power spectrum in contrast with cosmological observations.

\acknowledgments

We thank F. Arroja,   N. Bartolo, P. Creminelli, S. Matarrese, A. Ricciardone,   M. Sasaki and S. Yokoyama
for useful discussions and comments.   H. F. would like to thank ICTP and  University of Padova for hospitalities during the progress of this work and KITPC  for the hospitality  during the workshop `` Cosmology after PLANCK''  when this work was in progress.
R.E. thanks ICTP for the hospitality and for the support under the ``Sandwich Training Educational Program'' (STEP) fellowship.

\appendix{}

\section{The $\delta N$ analysis}
\label{deltaN-app}

In this Appendix we outline the details of the $\delta N$ analysis leading to Eqs.
(\ref{N-A})-(\ref{N-dotA-phi}). The starting equations are
\ba
\label{N-phi-Ab}
N &&= \frac{p_c}{2 \left( I -1 \right) } \ln \left( \frac{\phi}{\phi_{f}} \right)  \\
\label{N-phi-A2b}
&& = \frac{1}{3} \ln \left(\sqrt{\frac{3}{ 2 R} }  \left( \frac{ A -  A_{f} }{M_{P}} \right) + 1 \right) \, .
\ea
In addition during inflation, from the attractor condition we have
\ba
\label{R-defb}
R \equiv \frac{\dot A_x^2 f(\phi)^2 e^{-2 \alpha}}{2 V}  \simeq \frac{I \epsilon}{2}
\ea
in which the last approximate equality is for the general case when $p>p_c$ and $I \neq 0$.
In this view $R$ is treated as a free parameter which should be varied when calculating $\delta N$. This is because in general $N$ is defined in the phase space as a function of $(\phi, \dot \phi)$ and $(A_x, \dot A_x)$. Because of the  slow-roll conditions we can solve for $\dot \phi$ in terms of $\phi$ so we do not need to vary $\delta \dot \phi$ as an independent variable. However, for the gauge field, because of the  gauge invariance,
it is $\dot A_x$ (i.e. $F_{0 x}$) which is physical and not $A_x$ itself. The contribution of
gauge field $A_x$ (and not its derivative) only appears at the surface of end of inflation in which the gauge symmetry is spontaneously broken due to Higgs mechanism and the gauge field  becomes massive.  Therefore, $\delta N$ has contributions both from $\delta \dot A_x$
and $\delta A_x$.

Finally, the surface of end of inflation is parameterized by the angle $\gamma$ via
\ba
\label{phif-b}
\phi_f = \phi_c \cos \gamma \quad , \quad
A_f = \frac{g \phi_c}{\e} \sin \gamma \, .
\ea
Expressing the surface of end of inflation in this way, the dependent variables $\phi_f$ and
$A_f$ are traded in terms of the independent variable $\gamma$.

Now our goal is to use the constraint Eqs. (\ref{R-defb}) and (\ref{phif-b}) to find $\delta N$ in terns of initial fluctuations $\delta \phi, \delta A$ and $\delta \dot A$. In the expressions below, we keep $N$ general, but it is understood that we evaluate the initial perturbations
at the time of horizon crossing corresponding to $N=-60$.

Varying Eq. (\ref{N-phi-Ab}) with respect to variables $\phi, \phi_f$ and $I$ (or $R$)  yields
\ba
\label{deltaN-phi-2}
\delta N = N_{, X_I} \delta {X_I} + \frac{1}{2} N_{, X_I X_J} \delta X_I \delta X_J
\ea
in which $X_I$ collectively  represents the variables $\{ \phi, \phi_f, I \}$. Note that  when $I \neq 0$ and $R \simeq I \epsilon/2$, we can use $\delta I$ and $\delta R$
interchangeably. Calculating the derivatives, we have
\ba
N_{,\phi} &=& \frac{p_c }{2(I-1) \phi}   \quad , \quad  N_{,\phi_f} = \frac{-p_c }{2(I-1) \phi_f}
\quad , \quad
N_{, I} = -\frac{p_c}{2 (I-1)^2} \ln \frac{\phi}{\phi_f} = -\frac{N}{I-1}  \nonumber\\
N_{,\phi \phi } &=& -\frac{p_c }{2(I-1) \phi^2}   \quad , \quad
N_{,\phi_f \phi_f } = \frac{p_c }{2(I-1) \phi_f^2}   \quad , \quad
N_{, I I } = \frac{2 N}{(I-1 )^2} \nonumber\\
N_{,I \phi } &=&   -\frac{p_c}{2 (I-1)^2 \phi}  \quad , \quad
N_{,I \phi_f } =   \frac{p_c}{2 (I-1)^2 \phi_f}  \quad , \quad
N_{,\phi \phi_f } =0 \, .
\ea
Varying the attractor condition, Eq. (\ref{R-defb}), yields
\begin{align}
\label{attractor-per}
\frac{\delta R}{R} = \frac{\delta I}{I}&=  \frac{2 f_{,\phi}}{f}\delta \phi +  \frac{2 \delta \dot{A_x}}{\dot{A}} + \left( \frac{f_{,\phi \phi}}{f} + \bigg{(}\frac{f_{,\phi}}{f}\bigg{)}^2    \right)\delta \phi^2
+ \bigg{(}\frac{\delta \dot{A}}{\dot{A}}\bigg{)}^2 +  \frac{4 f_{,\phi}}{f} \frac{\delta \dot{A_x}}{\dot{A}} \delta \phi  \nonumber\\
& -2 \delta N \bigg{(}1+  \frac{2 f_{,\phi}}{f}\delta \phi +  \frac{2 \delta \dot{A_x}}{\dot{A}}   \bigg{)} + 2 \delta N^2
\end{align}
On the other hand, varying Eq. (\ref{phif-b}) with respect to the independent variable $\gamma$ to second order yields
\ba
\label{phif-gamma}
\delta \phi_f = \phi_c \left[ - \sin \gamma \left( \delta_1 \gamma + \delta_2 \gamma \right)
-\frac{1}{2} \cos \gamma \left( \delta_1 \gamma \right)^2 \right] \, ,
\ea
in which $\delta_1\gamma$ and $\delta_2\gamma$ respectively show the first and second order perturbations in $\gamma$.

Plugging Eqs. (\ref{phif-gamma}) and (\ref{attractor-per}) in Eq. (\ref{deltaN-phi-2}), and
neglecting the sub-leading terms $\delta N^2$ and $N I^2$ for $I \ll 1$,  yields
\ba
\label{deltaN-phi-a}
\delta N &\simeq& -\left[  \frac{p_c }{2 \phi} + 2 N I \frac{f_{, \phi}}{f}  \right] \delta \phi
+ 2 N I \frac{\delta \dot A_x}{\dot A_x} -\frac{p_c}{2} \tan \gamma \,  \delta_1 \gamma
-\frac{p_c}{2} \tan \gamma \,  \delta_2 \gamma  \nonumber\\
&+& \left[ N I  \left( \frac{f_{,\phi \phi}}{f} + \left(\frac{f_{,\phi}}{f} \right)^2    \right) -  \frac{p_c I \, f_{, \phi}}{ \phi f} + \frac{p_c}{4 \phi^2} \right] \delta \phi^2
+ N I  \left( \frac{\delta \dot A}{\dot A} \right)^2
+\left[ 4 N I  \frac{f_{, \phi} }{f} - \frac{I p_c }{\phi} \right]  \frac{\delta \dot A_x}{\dot A} \delta \phi \nonumber\\
&-& \frac{p_c}{4} \frac{(\delta_1\gamma)^2}{\cos^2 \gamma } -I p_c \tan \gamma
 \left[ \frac{ f_{,\phi}}{f} \delta \phi + \frac{\delta \dot A_x}{\dot A_x}
\right] \delta_1 \gamma
\ea

Similarly, varying Eq. (\ref{N-phi-A2b}) to second order yields
\ba
\label{deltaN-phi-2b}
\delta N = N_{, Y_I} \delta {Y_I} + \frac{1}{2} N_{, Y_I Y_J} \delta Y_I \delta Y_J
\ea
in which $Y_I$ collectively  represents the variables $\{ \phi, A_f, R \}$. Note that in formula for $N$ given in Eq. (\ref{N-phi-A2b}) it is $R$ and not $I$which  appears. This is crucial for the critical case in which $I=0$ but $R$ is still a free parameter containing information for $\dot A_x$.   Calculating the derivatives, we have
\ba
N_{, A} &=& \frac{e^{-3 N}}{M_P \sqrt{6 R}} \quad , \quad
N_{, A_f} = - \frac{e^{-3 N}}{M_P \sqrt{6 R}} \quad , \quad
N_{, R} = -\frac{1}{6 R}  \left( 1- e^{-3 N}  \right) \nonumber\\
N_{, A A} &=&   N_{, A_f A_f}= -\frac{e^{-6 N}}{2 R M_P^2} \quad , \quad
N_{, R R} = \frac{1}{12 R^2} \left( 2- e^{-3 N} - e^{-6 N} \right) \nonumber\\
N_{, A A_f} &=& \frac{e^{-6 N}}{2 R M_P^2}  \quad , \quad
N_{, R A} = - N_{, R A_f}=  -\frac{e^{-6 N}}{2 M_P \sqrt{6 R^3}}
\ea
On the other hand
\ba
\label{Af-gamma}
\delta A_f = \frac{g \phi_c}{\e} \left[ \cos \gamma ( \delta_1 \gamma + \delta_2 \gamma)
- \frac{1}{2} \cos \gamma (\delta_1 \gamma)^2 \right]  \, .
\ea
Plugging Eqs. (\ref{Af-gamma}) and (\ref{attractor-per}) in Eq. (\ref{deltaN-phi-2b}) to leading order yields
\ba
\label{deltaN-A-a}
\delta N  &=&  -\frac{1}{3 }  \left( 1- e^{-3 N}  \right)  \left[ \frac{f_{, \phi}}{f} \delta \phi + \frac{\delta \dot A}{\dot A} \right] +
\frac{e^{-3 N}}{\sqrt{6 R}}\frac{\delta A}{M_P}
-\frac{e^{-3 N}}{ \sqrt{6 R}} \frac{g \phi_c}{\e M_P}  \cos \gamma  \,
( \delta_1 \gamma  + \delta_2 \gamma)  \nonumber\\
&-&\frac{e^{-6 N}}{4 R }\frac{\delta A^2}{M_P^2}
+ \left[  \frac{g \phi_c \sin \gamma}{2 \e M_P} \frac{e^{-3 N}}{\sqrt{6 R}}
- \frac{g^2 \phi_c^2 \cos^2 \gamma }{2 \e^2  M_P^2} \frac{e^{-6 N}}{2 R}
\right] (\delta_1 \gamma)^2 + \frac{g \phi_c \cos \gamma}{2 R \e M_P^2} e^{-6 N}
\delta_1 \gamma \delta A\nonumber\\
&-& \frac{1}{6 }  \left( 1- e^{-3 N}  \right) \left[   \left( \left(\frac{f_{,\phi}}{f}\right)^2 + \frac{f_{, \phi \phi}}{f} \right) \delta \phi^2 + \left( \frac{\delta \dot A}{\dot A} \right)^2 + 4 \frac{f_{, \phi}}{f} \frac{\delta \dot A}{\dot A} \delta \phi \right] + \frac{1}{6}  \left( 2- e^{-3 N} - e^{-6 N} \right)
\left( \frac{f_{, \phi}}{f} \delta \phi + \frac{\delta \dot A}{\dot A}   \right)^2 \nonumber\\
&+& \frac{e^{-6 N}}{3 M_P} \sqrt{\frac{3}{2 R}}  \left( \frac{f_{, \phi}}{f} \delta \phi + \frac{\delta \dot A}{\dot A}   \right) \left( - \delta A   + \frac{g \cos \gamma}{\e} \delta_1 \gamma
\right)
\ea

Now we have two formulas for $\delta N$, given by Eqs. (\ref{deltaN-A-a}) and  (\ref{deltaN-phi-a}). Similar to \cite{Sasaki:2008uc} we can use these two equations to eliminate
$\delta_1 \gamma$ and $\delta_2 \gamma$ in terms of initial perturbations $\delta \phi, \delta A$ and $\delta \dot A$.
We find
\ba
\label{gamma1}
\delta_1 \gamma &=& \frac{\e}{g \phi_c  \cos \gamma} \left( \delta A
+ \sqrt{\frac{2 R}{3}} M_P \left(  \frac{\delta \dot A_x}{\dot A} + \frac{f_{, \phi}}{f} \delta \phi \right)  \right) \nonumber\\
\delta_2 \gamma &=&  \frac{\e M_P \sqrt{6 R}}{g \cos \gamma \, \phi_c}
\left[ \frac{\e \sin \gamma }{2g \cos^2 \gamma \, \phi_c }  \left( \frac{\delta A^2}{M_P \sqrt{6 R}}  + \frac{M_P}{3} \sqrt{\frac{2R}{3}} \left( \frac{\delta \dot A}{\dot A} \right)^2
+ \frac{2  \delta A . \delta \dot A}{3 \dot A}
+ \frac{2 f_{, \phi}}{3 f} \delta \phi \delta A \right) \right.  \nonumber\\
&+&  \left. \left( \frac{f_{, \phi \phi}}{6 f}
   + \frac{\e M_P \sin \gamma }{18 g \cos^2 \gamma \, \phi_c } \sqrt{6 R}
\left( \frac{f_{, \phi}}{f} \right)^2  \right) \delta \phi^2
+ \frac{ f_{, \phi}}{3 f} \left(1+  \frac{\e \sin \gamma M_P }{3 g \cos^2 \gamma \, \phi_c } \sqrt{6 R} \right) \delta \phi \frac{\delta \dot A}{\dot A}
\right]
\ea
Now we plug  back $\delta \gamma$ to either of Eqs. (\ref{deltaN-A-a}) or  (\ref{deltaN-phi-a}) yielding  our final formula for $\delta N$ given in Eqs.
(\ref{N-phi})-(\ref{N-dotA-phi}).


\section{Investigating the Validity of Separate Universe Assumption}
\label{separate-universe}

Before using the $\delta N$ formalism, one has to verify the validity of the
gradient expansion and the separate universes approach in our model. For isotropic FRW background containing interacting scalar fields this was studied in \cite{ Lyth:2004gb,  Naruko:2012fe, Naruko:2012um, Sugiyama:2012tj}. In this picture, each local Hubble patch behaves as a background FRW universe with the effects of a very long mode to rescale the background quantities such as $a(t), H, \rho $ and $p$ appropriately.
For the model of anisotropic inflation this was first demonstrated in \cite{Abolhasani:2013zya}. In this Appendix, we prove  the validity of separate universe approach in our current model of gauged hybrid inflation. Because of  the complexity of waterfall dynamics, we demonstrate this up to second order in perturbation theory which enables us to calculate the  power spectrum and the bispectrum. In principle, one can prove the validity of  $\delta N$ to all orders in perturbation theory, but this is beyond the scope of this work.

Let us start with the general Bianchi I background with three different background scale factors
\ba
\label{Bianchi-metric1}
ds^2 = -dt^2 + a_1(t)^2 d x^2 +a_2(t)^2 d y^2 +a_3(t)^2 d z^2 \, .
\ea
Following the notations used in \cite{Miedema:1993} we define
\ba
H_i(t)= \dfrac{\dot{a_i}}{a_i} \qquad , \qquad H \equiv \dfrac{1}{3} \sum_{i=1}^3 H_i \, .
\ea
Here $H_i$ represents the Hubble expansion rate for the $i$-th  direction with  $i=1,2,3$ and a dot indicates the derivative with respect to cosmic $t$.

To solve the background fields  equations one has  to specify the form of  energy momentum tensor. The general  form of energy momentum tensor $T_{\mu \nu}$ for an imperfect fluid is given by  \cite{ellis98}
\ba
\label{eq:stress}
T_{\mu \nu} = (\rho + p)\,u_{\mu}\,u_{\nu}+ p\,g_{\mu \nu}+ q_{\mu}\,u_{\nu} + u_{\mu}\,q_{\nu} +  \pi_{\mu \nu}
\ea
with the  conditions
\ba
 q_{\mu}\,u^{\mu} = 0 \quad , \quad   \pi^{\mu}{}_{\mu} = 0  \quad , \quad  ~\pi_{\mu \nu} = \pi_{\nu \mu} \quad , \quad
~\pi_{\mu \nu}\,u^{\nu} = 0 \, . \nonumber
\ea
Here $u^\mu$ represents  the four-vector associated with the fluid,  $\rho$ is the  energy density, $p $ represents  the isotropic pressure, $\pi_{\mu \nu} $ stands for the trace-free  anisotropic pressure (stress) and $q^\mu$  is the  heat conduction.

The background Einstein equations are
\ba
\label{00-back}
3  {\cal H}^2&\equiv& \sum_{i > j} \bar H_j \bar H_{j} =  \frac{\bar \rho}{M_P^2}
\\
\bar T^0{}_i &=& \bar q_i =0
\\
\label{i=j-back}
M_P^2 \dot{\bar H}_i&=&-3  M_P^2 \bar H \bar H_i + \dfrac{1}{2} (\bar \rho-\bar p) + \bar \pi^{i}{}_{i} \, .
\ea
In this notation, $\bar H_i$ represents the background Hubble expansion rates while  $\bar \rho, \bar p$ and  so on are the background values of the fluid's physical parameters.  Also note that we have defined ${\cal H}$ as the effective Hubble expansion rate appearing in Friedmann equation, Eq. (\ref{00-back}), which should not be mistaken with the conformal Hubble expansion rate usually used in literature.

On the other hand,  the energy conservation equation $u_{\mu} \nabla_{\nu} T^{\mu \nu} = 0$ yields
\ba
\label{cont.-back}
-u_{\mu} \nabla_{\nu} T^{\mu \nu} = \dot{\bar \rho} + 3 H (\bar \rho + \bar p) + \bar H_j \bar \pi
^{i}{}_{j} \delta^{j}_i=0 .
\ea
where  $\bar H = \sum_{i} \bar H_i/3$.

Let us now look at the perturbations.  We follow the notation used in \cite{Sugiyama:2012tj}
in our $\delta N$ analysis.  We denote the order of spatial derivative or the gradient expansion  by $\epsilon=k/aH$ while the  perturbations are denoted by $\delta$. In general, one has to allow three different gradient expansion parameters $\epsilon_i$ for three different spatial directions $\epsilon_i = k/a_i H_i$. However, in order to simplify the analysis we assume $\epsilon_i \sim \epsilon$ but the extension  to the general case will be easy.

Using the standard ADM formalism the perturbations in metric are parameterized as
\ba
\label{ADM}
ds^2 = -{\cal N}^2 dt^2 + \gamma_{ij} \left( dx^i + \beta^i dt \right) \left( dx^j + \beta^j dt\right) \, ,
\ea
in which, as usual, ${\cal N}$ is the lapse function, $\beta_i$ represents the shift vectors and $\gamma_{ij}$ is the three-dimensional  spatial metric.  Furthermore,  it is instructive to decompose the spatial metric into
\ba
\label{gamma-ij}
\gamma_{ij} = a_i(t) a_j(t) e^{\psi_i (\mathbf{x},t)+\psi_j (\mathbf{x},t)} \tilde{\gamma}_{ij} \, ,
\ea
in which  $a_i(t)$ represents  the scale factor for the $i$-th spatial direction and $\psi_i(\mathbf{x},t)$ is similar  to curvature perturbation $\psi$ for the isotropic background.

As studied in \cite{Abolhasani:2013zya},  one important step in  the analysis of the gradient expansion for the  Einstein equations  is the ordering of the shift functions $\beta^i$. In \cite{Abolhasani:2013zya}, for the model of anisotropic inflation, it was shown that $\beta^{i} = {\cal O} (\epsilon)$ to all orders in perturbation theory.  Here we demonstrate that, although quantum back-reactions induce non-zero shift function  $\beta^i$, these corrections can still be neglected as the shift function is at the second order in perturbation theory $\beta^{i} = {\cal O} (\delta^2)$. In the following we demonstrate this point in details.

In order to  find an estimation for the shift function $\beta^{i}$, it is enough to use  $ (0,i)$ component of the  Einstein equation. Using the definition of energy-momentum tensor for an imperfect fluid, one  finds
\ba
\delta T^i_0  = -{\cal N} \delta q^i + \epsilon {\cal O} (\delta) + \beta {\cal O} (\delta) \, .
\ea
On the other hand, employing the $ (i,0)$ component of Einstein equation, the heat transfer can be related to the shift function as
\ba
\delta q^i = - (\bar{\rho}+\bar{p}+\bar{\pi}^i_i) \beta^i + \epsilon {\cal O} (\delta) + \beta {\cal O} (\delta).
\ea
 Let us simplify the above equation in order to estimate the  amplitude of $\beta^i$. In our specific model one can simply show that $\bar{\pi}^i_i \sim I H^2 \ll \epsilon H^2$. As a result the shift function $\beta^i$ can be estimated as
\ba
\label{beta-i}
\beta^i \simeq \dfrac{-\delta  q^i}{ \epsilon H^2 M^2_{P}},
\ea
in which $\epsilon$ denotes the slow-roll parameter and should not be mistaken with the gradient expansion parameter. Now using the action, Eq. (\ref{action3}), the heat transfer can be read as
\ba
\label{heat-transfer}
\delta q^i = - \dfrac{1}{\cal N} \left[f^2(\phi) T^i _m {}^{em} + i e \left(\delta \psi^{\ast} \partial_0 \delta \psi -\delta \psi \partial_0 \delta \psi ^{\ast}\right) A^i + e^2 \delta \psi^2 A_0 A^i  \right] \, .
\ea
It has been previously shown in \cite{Abolhasani:2013zya} that the first term in the big bracket above, denoting the heat transfer of electromagnetism, is ${\cal O} (\epsilon)$. Below we find the order of  the gradient expansion of the  other two terms in
Eq. (\ref{heat-transfer}).

In the main text we chose unitary gauge by setting the complex phase of the waterfall field to be zero. In the following, it is more helpful to choose the Coulumb-radiation gauge in which  $A_0 =\partial_iA_i=0$. As a result, the third term in Eq. (\ref{heat-transfer}) is zero and we are left only with the second term in Eq. (\ref{heat-transfer}).

Note that the waterfall is extremely heavy during inflation so the background value of the waterfall field is pinned to zero before the waterfall phase transition. In addition, the quantum fluctuations of the waterfall field after horizon crossing are continuously  damped  till  the  time of waterfall instability  after which the modes start growing exponentially as a result of the tachyonic instability \cite{Lyth:2010ch, Abolhasani:2010kr, Abolhasani:2010kn, Abolhasani:2011yp, Fonseca:2010nk,
Gong:2010zf, Lyth:2012yp}.

Before doing any explicit calculation it is worth discussing how the waterfall dynamics can contribute to the heat transfer. For a moment suppose that every  waterfall quantum fluctuations  in the vicinity of the transition point has the following solution
\ba
\delta \psi_k(n)  = \delta \psi_k(0) ~ e^{-i \omega(t)\, t + \Omega(t)\, t} \, ,
\ea
in which $n\equiv N-N_c$  and $N_c$ represents the time of the waterfall phase transition.  Here $\omega(t)$ and $\Omega(t)$ are two real functions quantifying the frequency of the oscillations and the growth rate of each  mode respectively.

Looking at  Eq. \eqref{heat-transfer}, we conclude  that only the  oscillatory phase $\omega(t)$ of the solution can contribute into the second term in Eq. \eqref{heat-transfer}.  Therefore, only  modes which become classical but still have oscillatory behavior will contribute to this term. This takes place near the waterfall transition point when the modes   become classical but still may have weak oscillatory behaviors.

Let us now examine the above intuition more carefully. The evolution equation for the quantum mode $v_k \equiv a \delta \psi_k$ can be read as \cite{Abolhasani:2010kr}
\ba
v''_k+ \left(k^2 -\dfrac{2+\epsilon_{\psi}^2 n}{\tau^2} \right)v_k=0 \, ,
\ea
in which the prime denotes the derivation respect to conformal time $\tau$ and $\epsilon_\psi$
is a large number measuring the tachyonic mass of the waterfall quantum fluctuations \cite{Abolhasani:2010kr}.  In order to estimate the heat transfer one can assume that every mode becomes classical when $\omega_k \tau \sim \lambda \sim 1$ in  which $\omega_k$ is the time-dependent frequency  in the above equation $\omega_k^2 \equiv k^2 -(2+\epsilon_{\psi}^2 n)/\tau^2$. It is  evident  that there is a narrow band of momenta which can contribute to the heat transfer
\ba
\label{band}
k^2_{min} = \dfrac{\lambda+2+\epsilon_{\psi}^2}{\tau^2}<k^2<k^2_{max} = \dfrac{2+\epsilon_{\psi}^2}{\tau^2}.
\ea
As a starting point let us estimate the background value of the heat transfer
\ba
\label{heat-transfer-back}
\delta \bar{q}^i = - \dfrac{1}{\cal N} ~ i \e Im\left[\delta \psi^{\ast} \partial_0 \delta \psi \right]_{0} \bar{A}^i  \, ,
\ea
in which the charge density of waterfall field can be estimated as
\ba
\label{charge-back}
i \e  \, Im\left[\delta \psi^{\ast} \partial_0 \delta \psi \right]_{0} \simeq  \dfrac{e}{a}  \int_{k_{min}}^{k_{max}} d^3k~ \omega_{k}(t) |\delta \psi_k|^2 \, .
\ea
For simplicity, in the vicinity of the  transition point,  for the narrow band of momenta given in Eq. (\ref{band}), the frequency of the modes can be estimated as $\omega_k(t) \sim 1/\tau$ so one has
\ba
\label{charge-back}
i \e \, Im\left[\delta \psi^{\ast} \partial_0 \delta \psi \right]_{0} \sim  \e H  \dfrac{H^2_0}{4 \pi^2}  \, ,
\ea
in which we have  used the fact that the short modes around the transition point have the following amplitude \cite{Gong:2010zf}
\ba
\delta \psi_S (n)= \dfrac{H_0}{\sqrt{2k}k_c}e^{-n} \, .
\ea
Adding up these results, from Eq. (\ref{beta-i}),  one  finds the following relation for the background value of shift function $\bar \beta^i$
\ba
\bar \beta^i \sim \dfrac{eHA^i}{4\pi^2 \epsilon M^2_{P}} = {\cal P_R} \dfrac{\e A^i}{H} \, .
\ea
As one can see, the shift function has a small value at the background level. Although its value can be ignored in the background equation,  but still it can give rise to complications in the perturbation equations. Therefore, it is helpful to estimate the amplitude of $\beta^i$ induced from the waterfall effects.  Using the relation $a^{-2}\dot{A} = 2R ~V$, with $R$ being the  ratio of the gauge field energy density to the total energy density, one  finds that
\ba
\dfrac{\e A_f}{H} \simeq 3R \ll1.
\ea
Moreover, one has ${\cal P_R} = {\cal O}(\delta^2)$. As a result one concludes  that
\ba
\label{beta-order}
\bar{\beta}^i \sim R~ \delta^2  \ll  \delta^2 \, .
\ea
This indicates that $\bar \beta^i$ is not larger than the second order in perturbations.  Let us now estimate the amplitude of perturbations in shift function,  $\beta^i_k$,  for each mode $k$. Following the same method as above one can simply find that
\ba
\label{beta-pert}
\beta_k \sim  R \left(\delta \psi^2 \right)_{k}
\ea

Now Let us look at the $\delta N$ prescription for this model with a background value of $\beta^i$. In an FRW universe  the  local Hubble expansion rate for each direction in the presence of perturbations  is defined as \cite{Abolhasani:2013zya}
 \ba
H_i(\mathbf{x},t) = \dfrac{\bar{H}_i + \dot{\psi} (\mathbf{x},t)}{{\cal N}}.
 \ea
With some analysis one can  can show that in the case of non-zero shift function $\beta^i$, the above prescription is modified to
\ba
H_i(\mathbf{x},t) = \dfrac{\bar{H}_i + \dot{\psi} (\mathbf{x},t)+\partial_i \beta^i + \sum_{i \neq j} \beta_j \partial^j \psi^i -\beta_i \partial^i \psi^i }{{\cal N}}.
\ea
with no sum on repeated  $i$ indices.

The  $\delta N$ formalism is at hand noting that from the  equations above one has the following formula for the number of e-fold expansion for each direction
\ba
\label{Ni-int}
N_i(\mathbf{x},t_1,t_2) \equiv \int_{t_1}^{t_2} H_i (\mathbf{x},t) {\cal N} dt = \int_{t_1}^{t_2} \bar H_i  dt + \int_{t_1}^{t_2} \dot{\psi}_i  dt + \int  \partial_i \beta^i dt + \int \left( \beta_j \partial^j \psi^i -\beta_i \partial^i \psi^i \right) dt.
\ea
The first two terms above are the same as in \cite{Abolhasani:2013zya} while the remaining terms originate from the presence  of non-zero $\beta^i$. With a simple reasoning one can show that these additional terms do not play any role for the correlation functions which we are interested in. From Eq. \eqref{beta-order}, one concludes that the contribution of the last term in Eq. (\ref{Ni-int}) is at the third order of perturbation theory while to calculate the  power spectrum and the bispectrum we need  $\delta N$ formula up to second order in perturbations.

Moreover, It can be shown that the contribution of  $\partial_i \beta^i$ in Eq. (\ref{Ni-int}) is  not important. To see this note that  any Fourier mode of the square of the waterfall perturbations $\left(\delta \psi^2 \right)_k$ can have non-zero correlations only with itself. In other words, the square of waterfall fluctuations $\left(\delta \psi^2 \right)_k$ can be treated as  an individual fluctuation similar to fluctuation of other primary fields such as $\delta \phi$. In this view, the only connected diagram associated with the two-point correlation function which contains $\left(\delta \psi \right)_k$ is $\langle  \left(\delta \psi^2 \right)_k~ \left(\delta \psi^2 \right)_{k'} \rangle $. It is vivid that this contribution to the two point correlation function is at the fourth order of perturbation theory while the leading order terms from other fields are at the  second order. Similarly, the contribution of the waterfall field in the bispectrum can just emerge from the contractions of  the form $\langle  \left(\delta \psi^2 \right)_k~ \left(\delta \psi^2 \right)_{k'}~ \left(\delta \psi^2 \right)_{k''} \rangle $ which is at the sixth order in perturbation theory, negligible compared to leading terms  which are at the third order.  Therefore, one  concludes that despite the presence of a non-zero shift function in the vicinity of waterfall transition, any corrections due to heat transfer modification are too small to disrupt the  separate Universe assumption.

Finally, from Eq. (\ref{Ni-int}) one concludes
\ba
N_i(\mathbf{x},t_1,t_2) - \bar{N}_i(t) = \psi_i(t_2) -  \psi_i(t_1) + {\cal O} (\epsilon^2, \delta^3) \, ,
\ea
which, up to  ${\cal O} (\epsilon^2, \delta^3)$,  is the same  $\delta N$ formula as in \cite{Abolhasani:2013zya}.  As it is shown in Appendix  \ref{deltaN-app}, to calculate the power spectrum and the bispectrum we need
$\delta N$ formula only up to second order in perturbations. Therefore,  we conclude  that it is legitimate to apply $\delta N$ formalism in our analysis of power spectrum and bispectrum.

Finally we comment that the waterfall dynamics is not expected to induce observable curvature perturbations on large scales as studied in  e.g. \cite{Abolhasani:2010kr, Abolhasani:2011yp,      Fonseca:2010nk, Gong:2010zf, Lyth:2010ch, Lyth:2012yp}, for related works see also
\cite{Abolhasani:2010kn,    Levasseur:2010rk, Martin:2011ib, Clesse:2010iz,  Bugaev:2011qt, Kodama:2011vs, Mulryne:2011ni, Abolhasani:2012px,   Barnaby:2006km,   Barnaby:2006cq}. The waterfall dynamics affect only small scales, modes which leave the horizon around the time of waterfall.



\end{document}